\documentclass[english,aps]{revtex4}
\usepackage[T1]{fontenc}
\usepackage[latin9]{inputenc}
\usepackage{amsmath}
\usepackage{graphicx}
\usepackage{amssymb}
\makeatletter
\@ifundefined{textcolor}{}
{%
 \definecolor{BLACK}{gray}{0}
 \definecolor{WHITE}{gray}{1}
 \definecolor{RED}{rgb}{1,0,0}
 \definecolor{GREEN}{rgb}{0,1,0}
 \definecolor{BLUE}{rgb}{0,0,1}
 \definecolor{CYAN}{cmyk}{1,0,0,0}
 \definecolor{MAGENTA}{cmyk}{0,1,0,0}
 \definecolor{YELLOW}{cmyk}{0,0,1,0}
 }

\usepackage{epstopdf}
\usepackage{hyperref}
\usepackage{amsfonts}
\makeatother
\usepackage{babel}

\begin{document}

\title{Typical-Medium Theory of Mott-Anderson Localization}

\author{V. Dobrosavljevi\'{c} }

\affiliation{Department of Physics and National High Magnetic Field Laboratory\\
Florida State University,Tallahassee, Florida 32310}
\begin{abstract}
The Mott and the Anderson routes to localization have long been recognized
as the two basic processes that can drive the metal-insulator transition
(MIT). Theories separately describing each of these mechanisms were
discussed long ago, but an accepted approach that can include both
has remained elusive. The lack of any obvious static symmetry distinguishing
the metal from the insulator poses another fundamental problem, since
an appropriate \emph{static} order parameter cannot be easily found.
More recent work, however, has revisited the original arguments of
Anderson and Mott, which stressed that the key diference between the
metal end the insulator lies in the dynamics of the electron. This
physical picture has suggested that the {}``typical'' (geometrically
averaged) escape rate $\tau_{typ}^{-1}=\exp\langle\ln\tau_{esc}^{-1}\rangle$
from a given lattice site should be regarded as the proper \emph{dynamical
order parameter} for the MIT, one that can naturally describe both
the Anderson and the Mott mechanism for localization. This article
provides an overview of the recent results obtained from the corresponding
\emph{Typical-Medium Theory}, which provided new insight into the
the two-fluid character of the Mott-Anderson transition. 
\end{abstract}

\maketitle

\section{From metal to insulator: a new perspective}

\textbf{Metal or insulator - and why?} To answer this simple question
has been the goal and the driving force for much of the physical science
as we know it today. Going back to Newton's not-so-successful exercises
in Alchemy, the scientist had tried to understand what controls the
flow of electricity in metals and what prevents it in insulators \cite{mott-book90}.
To understand it and to control it - achieving this could prove more
useful and lucrative than converting lead into gold. Indeed, the last
few decades have witnessed some most amazing and unexpected advances
in material science and technology. And this ability - its intellectual
underpinning - is what was indispensable in designing and fabricating
the iPhone, the X-Box, and the MRI diagnostic tool. Today's kids have
grown up in a different world than had their parents - all because
we have learned a few basic ideas and principles of electron dynamics.

In almost every instance, these advances are based on materials that
find themselves somewhere between metals and insulators. Material
properties are easy to tune in this regime, where several possible
ground states compete \cite{dagotto-2005-309}. Here most physical
quantities display unusual behavior \cite{RoP2005review}, and prove
difficult to interpret using conventional ideas and approaches. Over
the last few decades, scores of theoretical scenarios and physical
pictures have been proposed, most of which will undoubtedly end up
in back drawers of history. Last couple of years, however, have seen
a veritable avalanche of new experimental results, which provide compelling
clues as to what the theorists should not overlook: the significant
effects of spatial inhomogeneities in the midst of strongly correlated
phases.

\begin{figure}[h]
 \centering{}\includegraphics[width=12.5cm]{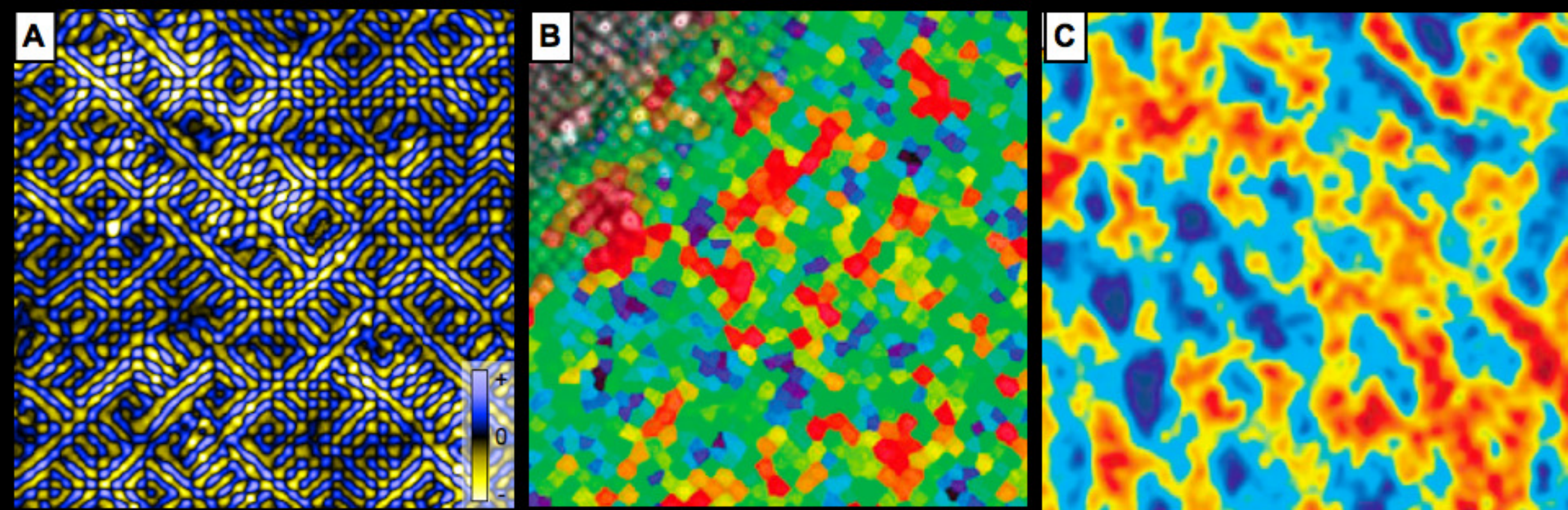}\vspace{12pt}

\caption{Spectacular advances in scanning tunneling microscopy (STM) have revealed
that many {}``bad metals'' or barely-doped insulators are surprisingly
inhomogeneous on the nano-scale. Understanding and controlling these
materials will not be possible without coming to grips with the origin,
the stability, and the statistics of such mesoscopic granularity.
(A) {}``Tunneling asymmetry'' imaging \cite{kohsaka-2007-315} provides
evidence for the emergence of a low temperature {}``electronic cluster
glass'' within the superconducting phase of $Ca_{1.88}Na_{0.12}CuO_{2}Cl_{2}$;
(B) Fourier-transform STM \cite{wise-2008} reveals nano-scale Fermi
surface variations in $Bi_{2}Sr_{2}CuO_{6-x}$; (C) Differential conductance
maps \cite{pasupathy-2008-320} showing spatial variations of the
local pseudogaps in the normal phase $(T\gg T_{c})$ of $Bi_{2}Sr_{2}CaCu_{2}O_{8+\delta}$. }

\end{figure}\pagebreak

To understand many, if not most exotic new materials, one has to tackle
the difficult problem of understanding the metal-insulator quantum
phase transition, as driven by the combined effects of strong electronic
correlations and disorder. Traditional approaches to the problem,
which emerged in the early 1980s, have focused on examining the perturbative
effects of disorder within the Fermi liquid framework. Despite their
mathematical elegance, these theories, unfortunately, prove ill-suited
to describe several key physical processes, such as tendency to local
magnetic moment formation and the approach to the Mott insulating
state. In addition, such weak-coupling theories cannot easily describe
strongly inhomogeneous phases, with behavior often dominated by broad
distributions and rare disorder configurations.

This new insight, which is largely driven by experimental advances,
seems to suggest that an alternative theoretical picture may provide
a better starting point. In this article we describe recent advances
based on a new theoretical method, which offers a complementary perspective
to the conventional weak-coupling theories. By revisiting the original
ides of Anderson and Mott, it examines the \emph{typical escape rate}
from a given site as the fundamental \emph{dynamical order parameter}
to distinguish between a metal and an insulator. This article describes
the corresponding \emph{Typical-Medium Theory (TMT)} and discusses
some of its recent results and potential applications. We fist discuss
discuss, in some detail, several experimental and theoretical clues suggesting
that a new theoretical paradigm is needed. The formulation of TMT
for Anderson localization of noninteracting electrons is then discussed,
with emphasis on available analytical results. Finally, we review
recent progress in applying TMT to the Mott-Anderson transition for
disordered Hubbard models, and discuss resulting the two-fluid behavior
at the critical point.

\section{Theoretical challenges: beyond Cinderella's slipper?}

The existence of a sharp metal-insulator transition at $T=0$ has
been appreciated for many years \cite{mott-book90}. Experiments on
many systems indeed have demonstrated that a well defined critical
carrier concentration can easily be identified. On the theoretical
side, ambiguities on how to describe or even think of the metal-insulator
transition have made it difficult to directly address the nature of
the critical region. In practice, one often employs the theoretical
tools that are available, even if possibly inappropriate. Even worse,
one often focuses on those systems and phenomena that fit an available
theoretical mold, ignoring and brushing aside precisely those features
that seem difficult to understand. This {}``Cinderella's slipper''
approach is exactly what one should not do; unfortunately it happens
all too often. A cure is, of course, given by soberly confronting
the experimental reality: what seems paradoxical at first sight often
proves to be the first clue to the solution.

\subsection{Traditional approaches to disordered interacting electrons}

Most studies carried out over the last thirty years have focused on
the limit of weak disorder \cite{leeramakrishnan}, where considerable
progress has been achieved. Here, for non-interacting electrons the
conductance was found to acquire singular (diverging) corrections
in one and two dimensions, an effect known as {}``weak localization''\cite{gang4,leeramakrishnan}.
According to these predictions, for $d\le2$ the conductivity would
monotonically decrease as the temperature is lowered, and would ultimately
lead to an insulating state at $T=0$. Interestingly, similar behavior
was known in Heisenberg magnets \cite{wegner79,wegner80,re:Goldenfeld92},
where it resulted from $d=2$ being the {\em lower critical dimension}
for the problem. This analogy with conventional critical phenomena
was first emphasized by the {}``gang of four'' \cite{gang4} , as
well as Wegner \cite{wegner79,wegner80}, who proposed an approach
to the metal-insulator transition based on expanding around two dimensions.
For this purpose, an effective low energy description was constructed
\cite{wegner79,wegner80,efetov80}, which selects those processes
that give the leading corrections at weak disorder in and near two
dimensions. This {}``non-linear sigma model'' formulation \cite{wegner79,wegner80,efetov80}
was subsequently generalized to interacting electrons by Finkelshtein
\cite{fink-jetp83}, and studied using renormalization group methods
in $2+\varepsilon$ dimensions \cite{fink-jetp83,fink-jetp84,cclm-prb84}.
In recent years, the non-linear sigma model of disordered interacting
electrons has been extensively studied by several authors \cite{belitz-kirkpatrick-rmp94,Punnoose}.

While the sigma model approach presented considerable formal complexity,
its physical content proved - in fact - to be remarkably simple. As
emphasized by Castellani, Kotliar and Lee \cite{ckl}, one can think
of the sigma model of disordered interacting electrons as a low energy
{\em Fermi liquid} description of the system. Here, the low energy
excitations are viewed as a gas of diluted quasi-particles that, at
least for weak disorder, can be described by a small number of Fermi
liquid parameters such as the diffusion constant, the effective mass,
and the interaction amplitudes. In this approach, one investigates
the evolution of these Fermi liquid parameters as weak disorder is
introduced. The metal-insulator transition is then identified by the
{\em instability} of this Fermi liquid description, which in $d=2+\varepsilon$
dimensions happens at weak disorder, where controlled {\em perturbative}
calculations can be carried out.\pagebreak

Remarkably, by focusing on such a stability analysis of the metallic
state, one can develop a theory for the transition which does not
require an {\em order parameter} description, in contrast to the
standard approaches to critical phenomena \cite{re:Goldenfeld92}.
This is a crucial advantage of the sigma model approach, precisely
because of the ambiguities in defining an appropriate order parameter.
We should stress, however, that by construction, the sigma model focuses
on those physical processes that dominate the perturbative, weak disorder
regime. In real systems, the metal-insulator transition is found at
strong disorder, where a completely different set of processes could
be at play.

\subsection{Anderson's legacy: strong disorder fluctuations}

From a more general point of view, one may wonder how pronounced are
the effects of disorder on quantum phase transitions. Impurities and
defects are present in every sample, but their full impact has long
remained ill-understood. In early work, Griffiths discovered \cite{griffiths-jmp67,griffiths-prl69}
that rare events due to certain types of disorder can produce nonanalytic
corrections in thermodynamic response. Still, for classical models
and thermal phase transitions he considered, these effects are so
weak to remain unobservably small \cite{lohneysen-etal-rmp07}. The
critical behavior then remains essentially unmodified.

More recent efforts turned to quantum ($T=0$) phase transitions \cite{sachdev-book},
where the rare disorder configurations prove much more important.
In some systems they gives rise to {}``Quantum Griffiths Phases''
(QGP) \cite{RoP2005review,vojta-review06}, associated with the {}``Infinite
Randomness Fixed Point {}`` (IRFP) phenomenology\cite{fisher95}.
Here, disorder effects produce singular thermodynamic response not
only at the critical point, but over \textit{an entire region} in
its vicinity. In other cases, related disorder effects are predicted
\cite{vojta-prl03,hoyos-vojta-prl08} to result in {}``rounding''
of the critical point, or to produce intermediate {}``cluster glass''
phases \cite{dobrosavljevic-miranda-prl05,case-dobrosavljevic-prl07}
masking the critical point. Physically, QGP-IRFP behavior means \cite{RoP2005review,vojta-review06}
that very close to the critical point, the system looks increasingly
inhomogeneous even in static response.

But how robust and generic may such pronounced sensitivity to disorder
be in real systems? Does it apply only to (magnetic and/or charge)
ordering transitions, or is it relevant also for the metal-insulator
transitions (MITs)? A conclusive answer to these questions begs the
ability to locally visualise the system on the nano-scale. Remarkably,
very recent STM images provide striking evidence of dramatic spatial
inhomogeneities in surprisingly many systems. While much more careful
experimental and theoretical work is called for, these new insights
makes it abundantly clear that strong disorder effects - as first
emphasized by early seminal work of Anderson \cite{anderson58} -
simply cannot be disregarded.

\begin{figure}[h]
 
\begin{centering}
\includegraphics[width=12.5cm]{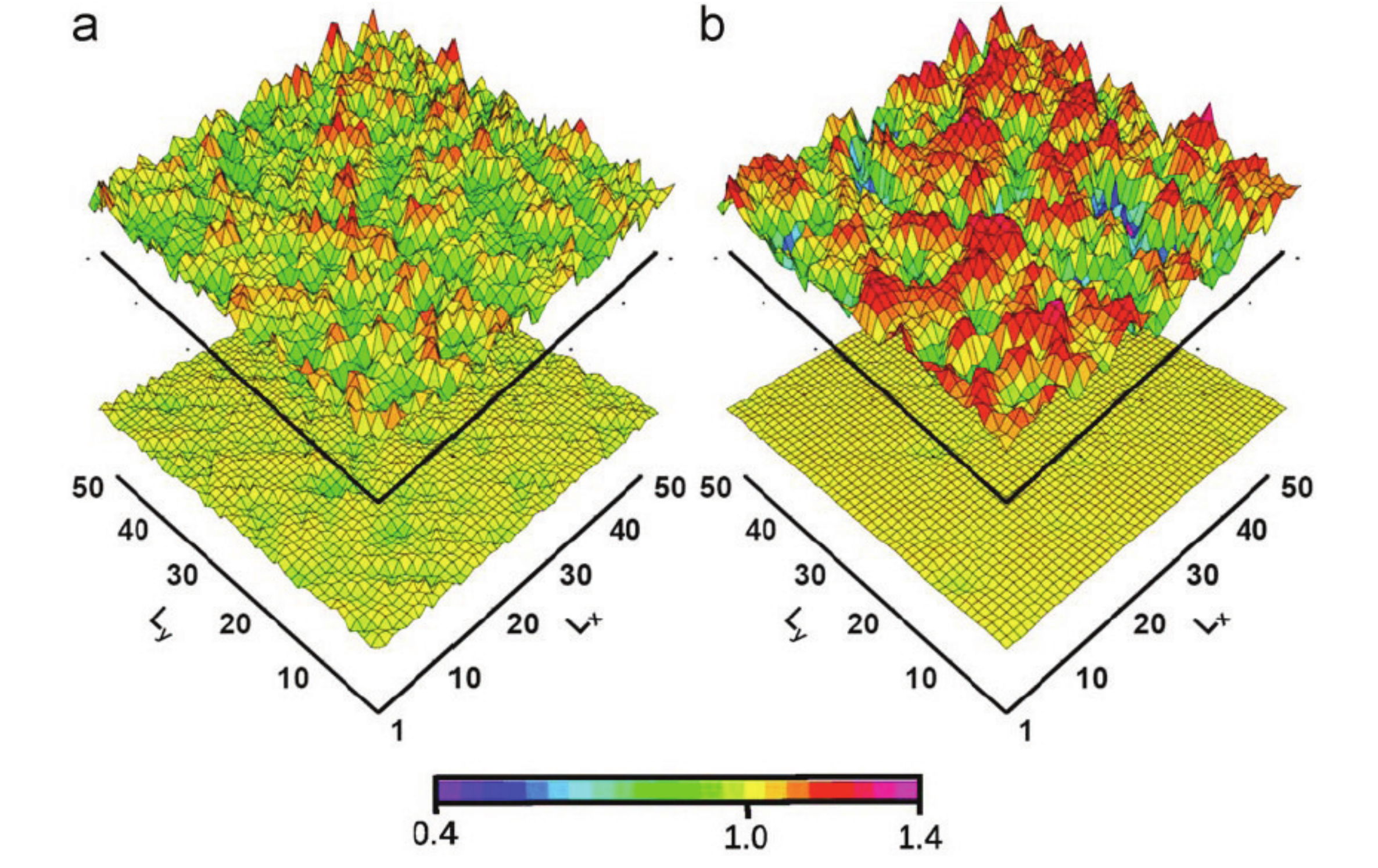} 
\par\end{centering}

\caption{Theory predictions \cite{andrade09physicsB}for an {}``Electronic
Griffiths Phase'' \cite{andrade09prl} in a moderately disordered
normal metal near a Mott meal-insulator transition. Local density
of states (LDOS) spectra look dramatically {}``smoother'' near the
Fermi energy (bottom) than away from it (top). This contrast is more
pronounced close to the Mott transition (right), than outside the
critical region (left). Very similar behavior was recently observed
by STM imaging of the superconducting phase of doped cuprates \cite{kohsaka-2007-315},
but our results strongly suggest that such energy-resolved {}``disorder
healing'' \cite{tanaskovicetal03,andrade09physicsB,andrade09prl}is
a much more general property of Mott systems.}

\end{figure}

\subsection{The curse of Mottness: the not-so-Fermi liquids }

One more issue poses a major theoretical challenge. According to Landau's
Fermi liquid theory, any low temperature metal behaves in a way very
similar to a gas of weakly interacting fermions. In strongly correlated
systems, closer to the Mott insulating state, this behavior is typically
observed only below a modest crossover temperature $T^{*}\ll T_{F}$.
Adding disorder typically reduces $T^{*}$ even further, and much
of the experimentally relevant temperature range simply does not conform
to Landau's predictions. Theoretically, this situation poses a serious
problem, since the excitations in this regime no longer assume the
character of diluted quasiparticles. Here perturbative corrections
to Fermi liquid theory simply do not work \cite{RoP2005review}, and
a conceptually new approach is needed. %
\begin{figure}[h]

\begin{centering}
\includegraphics[width=12.5cm]{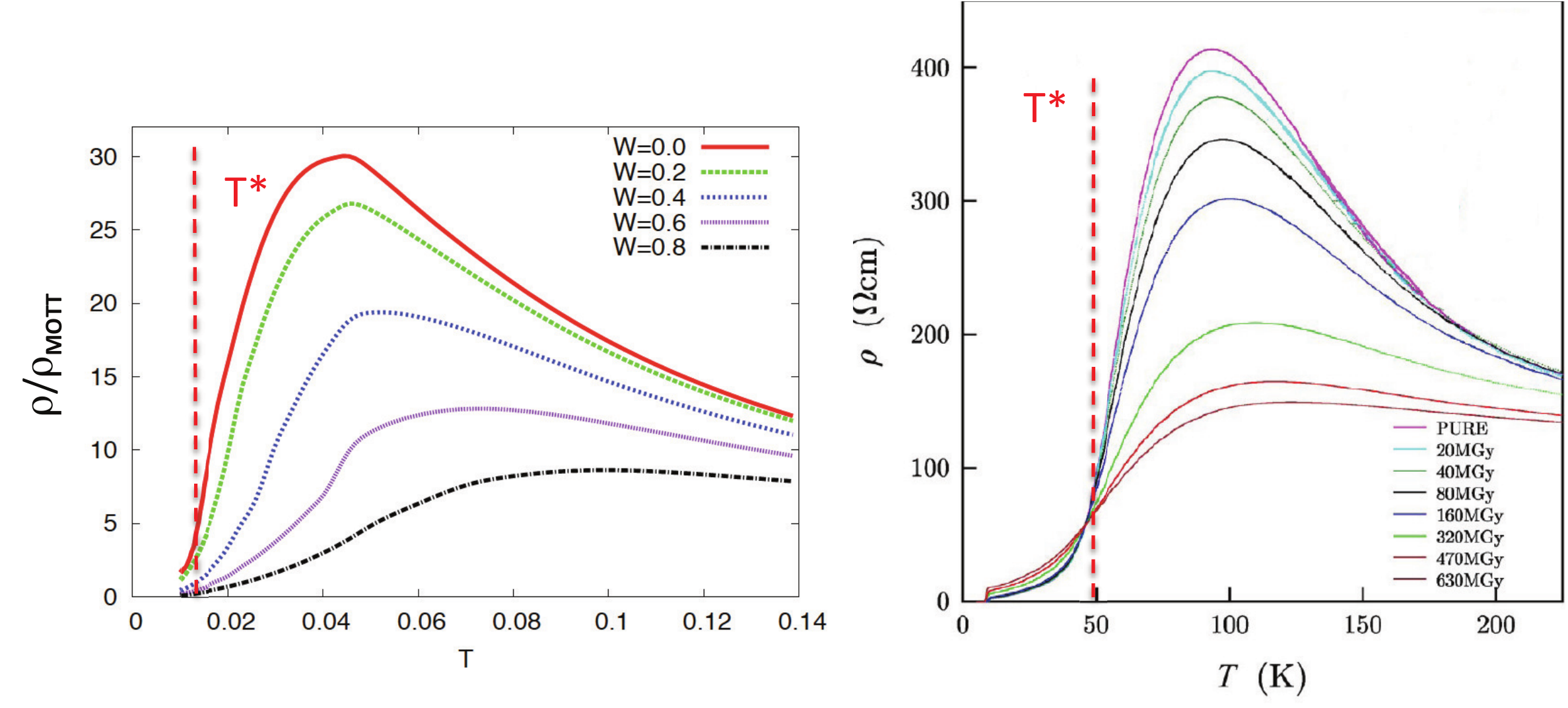} 
\par\end{centering}

\caption{Finite temperature metal-insulator crossover in transport close to
a disordered Mott transition. Very high values of resistivity, strongly
exceeding the {}``Mott limit'' \cite{mott-book90,hussey-2004-84}
are observed above the crossover temperature $T^{*}$. Remarkably,
increasing disorder $W$ reduces the resistivity maximum, rendering
the system effectively more metallic. This behavior, which is clearly
seen in our DMFT modelling \cite{radonjic10prb} (left panel), has
very recently been also observed in experiments \cite{analytis06prl}
on organic charge-transfer salts (right panel), where disorder is
systematically introduced by X-ray irradiation. }

\end{figure}

A new theoretical paradigm, which works best precisely in the incoherent
metallic regime, has been provided by the recently developed \emph{Dynamical
Mean-Field Theory} (DMFT) methods \cite{georgesrmp}. Unfortunately,
in its original formulation, which is strictly exact in the limit
of infinite dimensions, DMFT is not able to capture Anderson localization
effects. Over the last twelve years, this nonperturbative approach
has been further extended \cite{RoP2005review,vladtedgabi,dk-prb94,motand,vladgabisdmft2,tmt,aguiaretal1,aguiaretal2,mirandavlad3}
to incorporate the interplay between the two fundamental mechanisms
for electron localization: the Mott (interaction-driven) \cite{mott-book90}
and the Anderson (disorder-driven) \cite{anderson58,andersonlocrev}
route to arrest the electronic motion. In addition, the DMFT formulation
can be very naturally extended to also describe strongly inhomogeneous
and glassy phases of electrons \cite{pastor-prl99,mitglass-prl03,Denis,arrachea04prb,pankov05prl,muller04prl,re:Sachdev96,senguptageorges,westfahljr-2003-68,wu04prb},
and even capture some aspects of the Quantum Griffiths Phase physics
found at strong disorder \cite{RoP2005review,dkk,motand,vladgabisdmft2,mirandavladgabi1,mirandavladgabi2,mirandavladgabi3,mirandavlad1,mirandavlad2,mirandavlad3,tanaskovicetal04,tanaskovic-2005-95,dobrosavljevic-miranda-prl05,case-dobrosavljevic-prl07}.
In the following, we first discuss the DMFT method as a general order-parameter
theory for the metal-insulator transition, and the explain how it
needs to be modified to capture Anderson localization effects.

\section{Order-parameter approach to interaction-localisation}

\subsection{Need for an order-parameter theory: experimental clues}

In conventional critical phenomena, simple mean-field approaches such
as the Bragg-Williams theory of magnetism, or the Van der Waals theory
for liquids and gases work remarkably well - everywhere except in
a very arrow critical region. Here, effects of long wavelength fluctuations
emerge that modify the critical behavior, and its description requires
more sophisticated theoretical tools, based on renormalization group
(RG) methods. A basic question then emerges when looking at experiments:
is a given phenomenon a manifestation of some underlying mean-field
(local) physics, or is it dominated by long-distance correlations,
thus requiring an RG description? For conventional criticality the answer
is well know, but how about metal-insulator transitions? Here the
experimental evidence is much more limited, but we would like to emphasize
a few well-documented examples which stand out.

\subsubsection{Doped semiconductors}

Doped semiconductors such as Si:P \cite{doped-book} are the most
carefully studied examples of the MIT critical behavior. Here the
density-dependent conductivity extrapolated to $T=0$ shows sharp
critical behavior \cite{paalanen91} of the form $\sigma\sim(n-n_{c})^{\mu}$,
where the critical exponent $\mu\approx1/2$ for uncompensated samples
(half-filled impurity band), while dramatically different $\mu\approx1$
is found for heavily compensated samples of Si:P,B, or in presence
of strong magnetic fields. Most remarkably, the dramatically differences
between these cases is seen over an extremely broad concentration
range, roughly up to several times the critical density. Such robust
behavior, together with simple apparent values for the critical exponents,
seems reminiscent of standard mean-field behavior in ordinary criticality.%

\begin{figure}[h]

\begin{centering}
 \includegraphics[width=9cm]{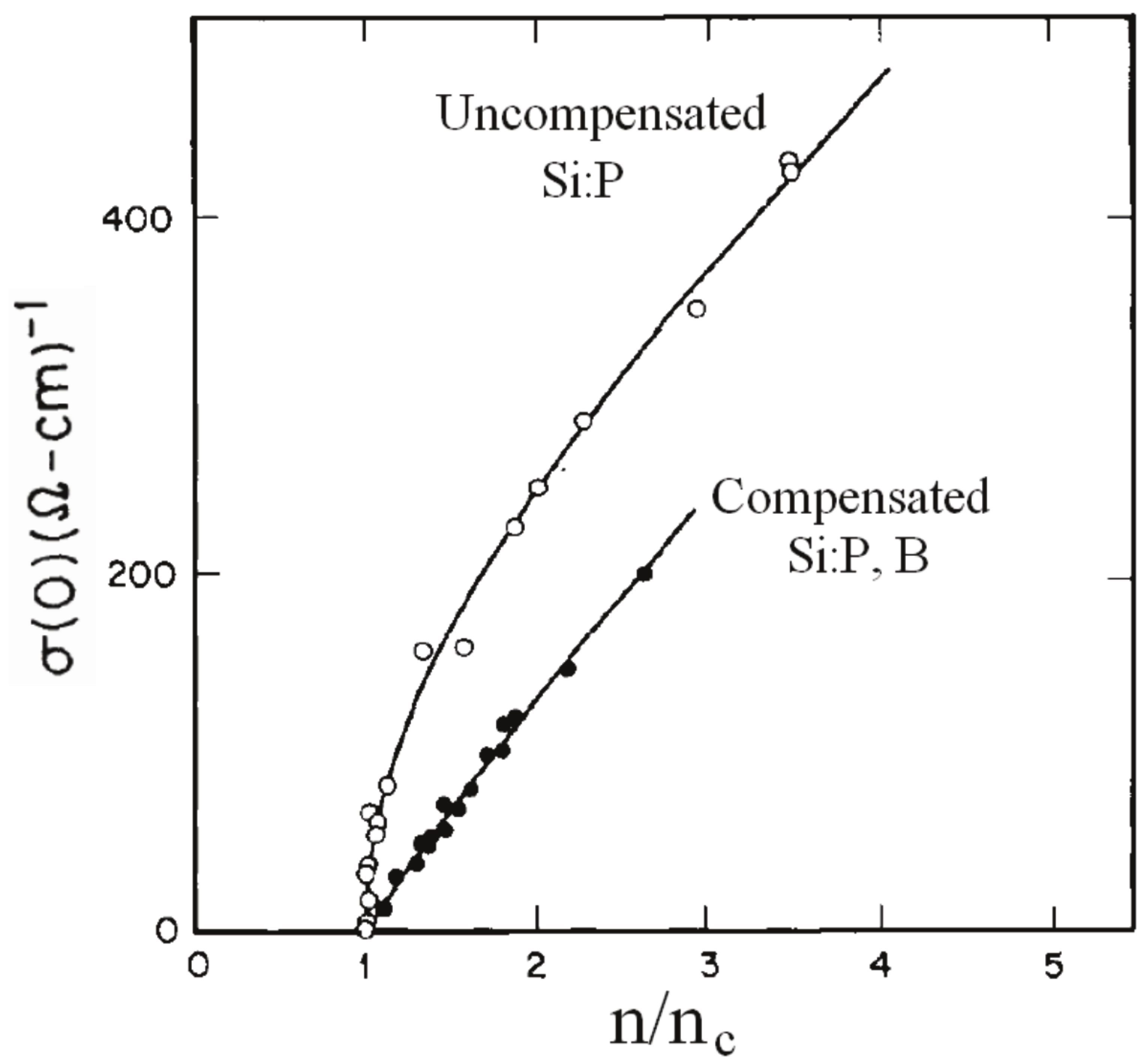}
 \par\end{centering}

\caption{Critical behavior of the conductivity for uncompensated Si:P and compensated
Si:P,B \cite{paalanen91}. The conductivity exponent $\mu\approx1/2$
in absence of compensation, while $\mu\approx1$ in its presence.
Clearly distinct behavior is observed in a surprisingly broad range
of densities, suggesting mean-field scaling. Since compensation essentially
corresponds to carrier doping away from a half-filled impurity band
\cite{doped-book}, it has been suggested \cite{leeramakrishnan}
that the difference between the two cases may reflect the role of
strong correlations.}

\end{figure}

\subsubsection{2D-MIT}

\begin{figure}
\includegraphics[width=6cm]{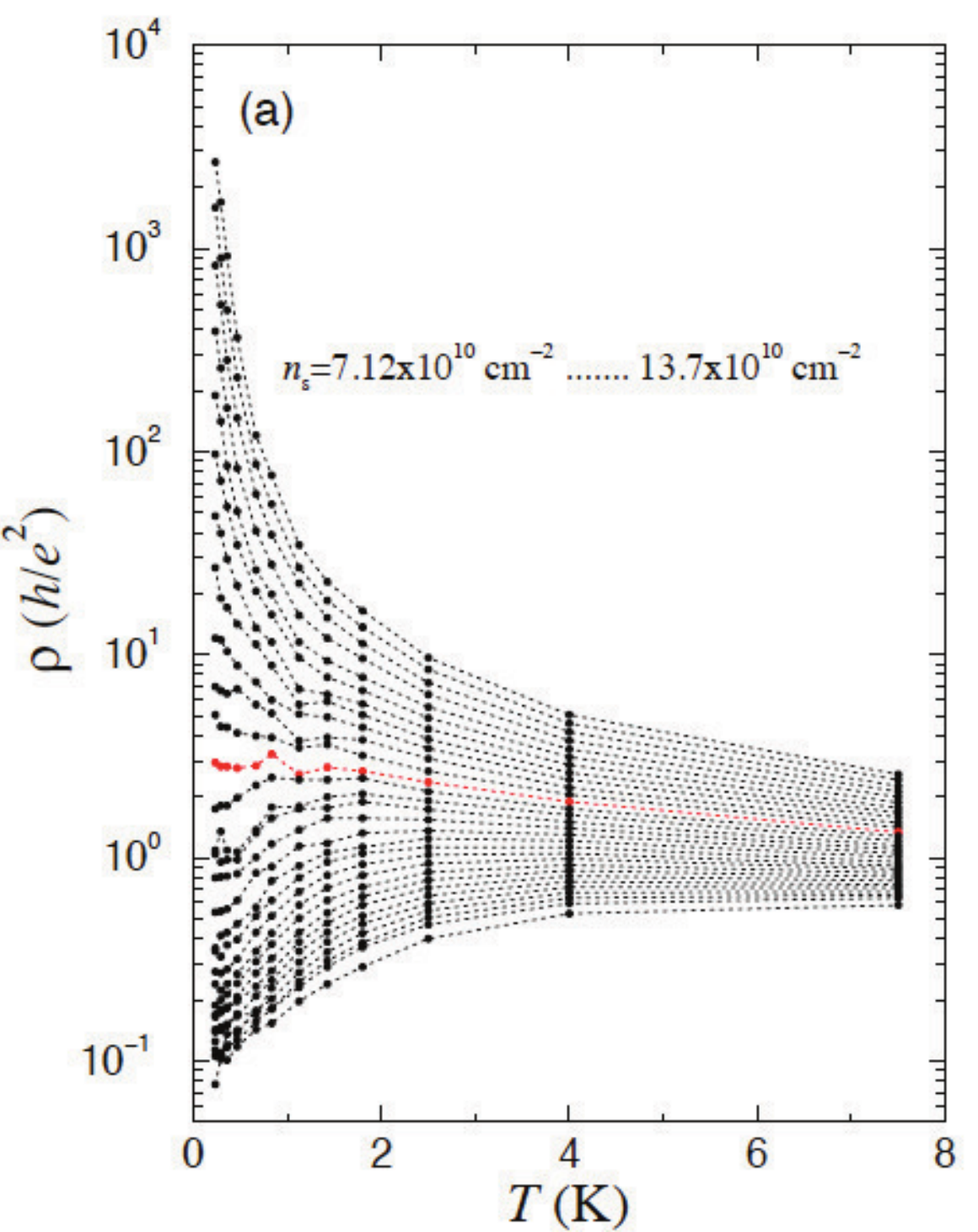}\includegraphics[width=6.2cm]{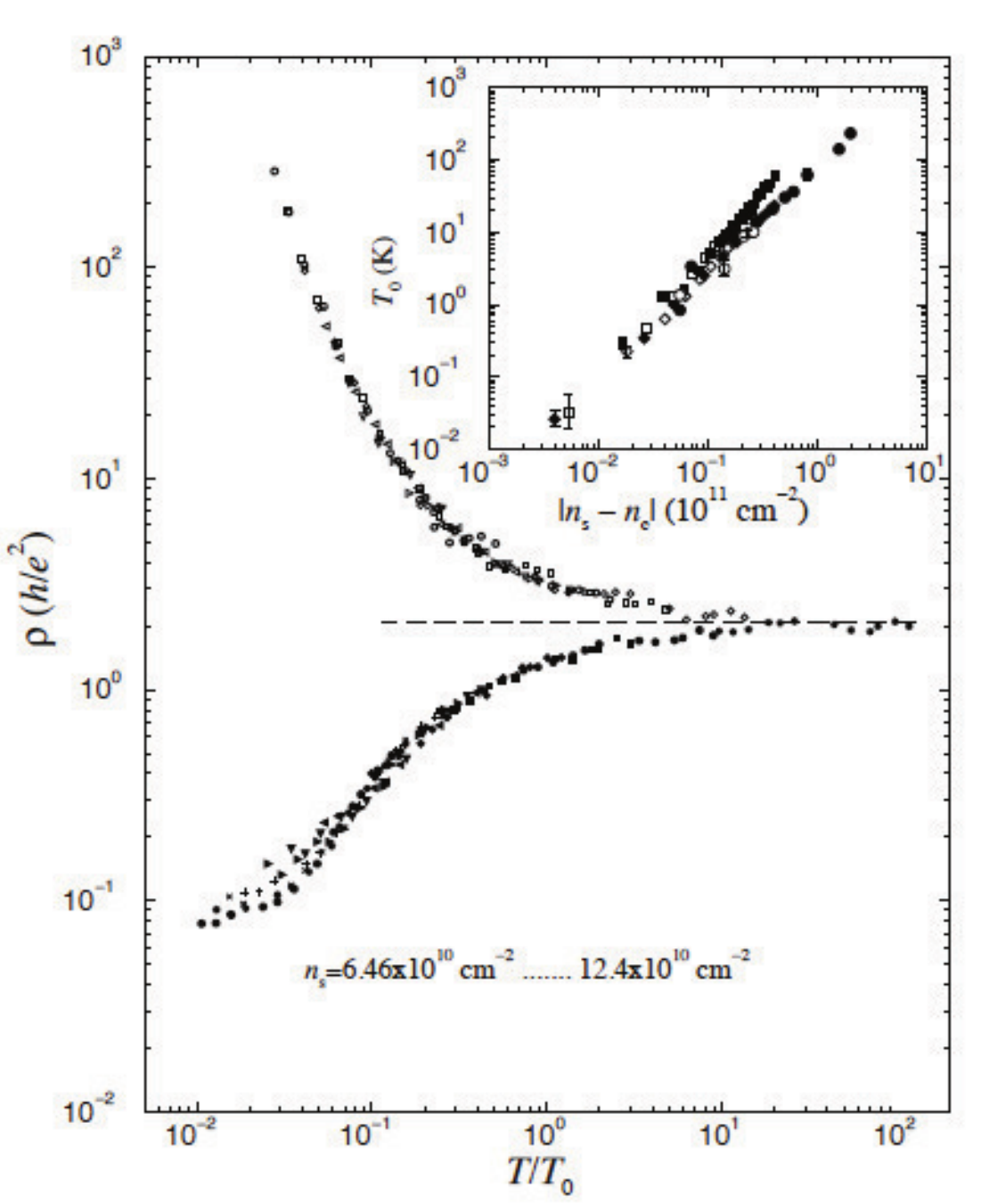}

\caption{The resistivity curves (left panel) for a two-dimensional electron
system in silicon \cite{re:Kravchenko95} show a dramatic metal-insulator
crossover as the density is reduced below $n_{c}\sim10^{11}cm^{-2}$.
Note that the system has {}``made up its mind'' whether to be a
metal or an insulator even at surprisingly high temperatures $T\sim T_{F}\approx10K$.
The right panel displays the scaling behavior which seems to hold
over a comparable temperature range. The remarkable {}``mirror symmetry''
\cite{simonian97prb} of the scaling curves seems to hold over more
then an order of magnitude for the resistivity ratio. This surprising
behavior has been interpreted \cite{gang4me} as evidence that the
transition region is dominated by strong coupling effects characterizing
the insulating phase. }

\end{figure}

Signatures of a remarkably sharp metal-insulator transition has also
been observed \cite{re:Kravchenko95,re:Popovic97,abrahams-rmp01}
in several examples of two-dimensional electron gases (2DEG) such
as silicon MOSFETs. While some controversy regarding the nature or
even the driving force for this transition remains a subject of intense
debate, several experimental features seem robust properties common
to most studied samples and materials. In particular, various experimental
groups have demonstrated \cite{re:Kravchenko95,re:Popovic97} striking
scaling of the resistivity curves in the critical region, which seems
to display \cite{simonian97prb} remarkable mirror symmetry ({}``duality'')
\cite{gang4me} over a surprisingly broad interval of parameters.
In addition, the characteristic behavior extends to remarkably high
temperatures, which are typically comparable the Fermi temperature
\cite{abrahams-rmp01}. One generally does not expect a Fermi liquid
picture of diluted quasiparticles to apply at such {}``high energies'',
or any correlation length associated with quantum criticality to remain
long. 

These experiments taken together provide strong hints that an appropriate
mean-field description is what is needed. It should provide the equivalent
of the Van der Waals equation of state, for disordered interacting electrons.
Such a theory has long been elusive, primarily due to a lack of a
simple order-parameter formulation for this problem. Very recently,
an alternative approach to the problem of disordered interacting electrons
has been formulated, based on dynamical mean-field (DMFT) methods
\cite{georgesrmp}. This formulation is largely complementary to the
scaling approach, and has already resulting in several striking predictions.
In the following, we briefly describe this method, and summarize the
main results that have been obtained so far.

\subsection{The DMFT physical picture}

The main idea of the DMFT approach is in principle very close to the
original Bragg-Williams (BW) mean-field theories of magnetism \cite{goldenfeldbook}.
It focuses on a single lattice site, but replaces \cite{georgesrmp}
its environment by a self-consistently determined {}``effective medium'',
as shown in Fig. 1.3.

\begin{figure}[h]
 \centerline{\includegraphics[width=10cm]{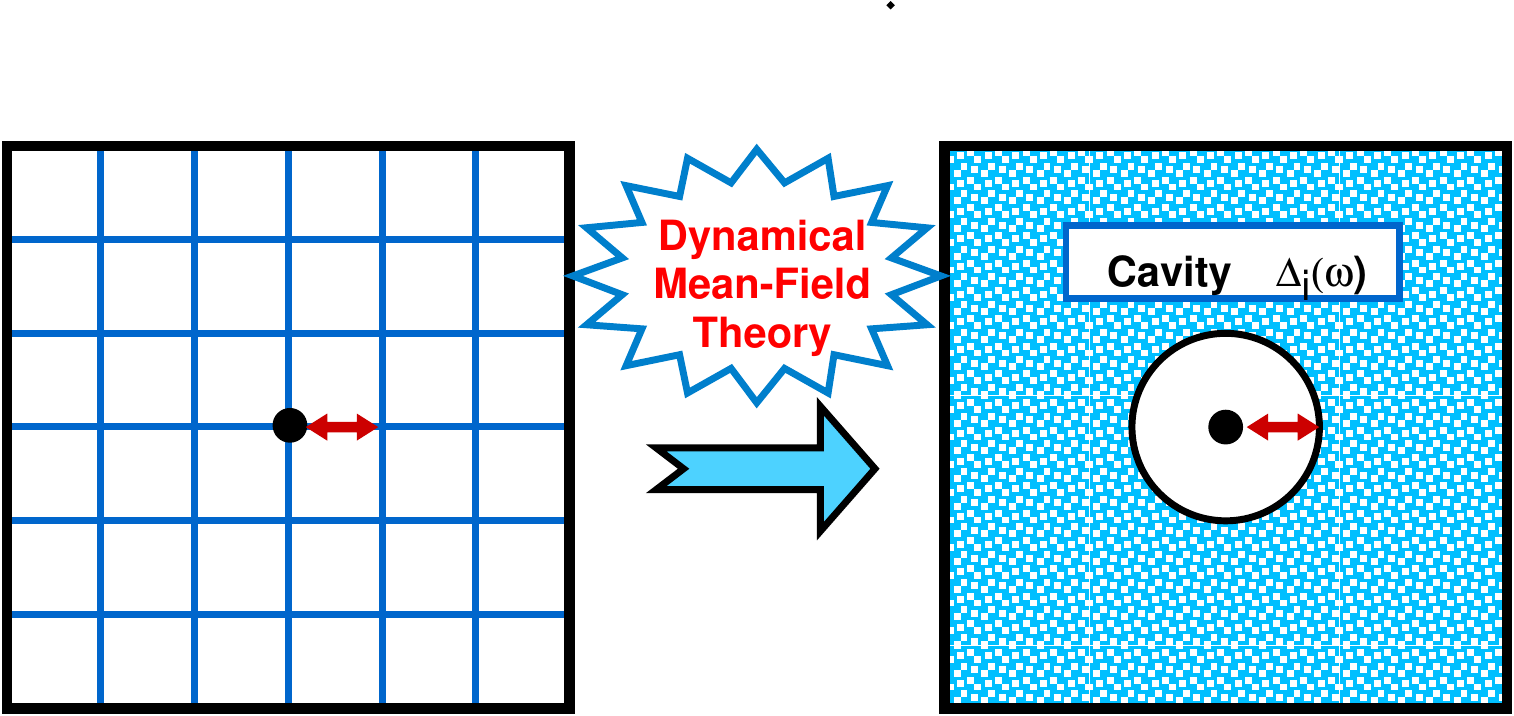}} \vspace{18pt}

\caption{In dynamical mean-field theory, the environment of a given site is
represented by an effective medium, represented by its {}``cavity
spectral function'' $\Delta_{i}(\omega)$. In a \emph{disordered}
system, $\Delta_{i}(\omega)$ for different sites can be very different,
reflecting Anderson localization effects.}

\end{figure}

In contrast to the BW theory, the environment cannot be represented
by a static external field, but instead must contain the information
about the dynamics of an electron moving in or out of the given site.
Such a description can be made precise by formally integrating out
\cite{georgesrmp} all the degrees of freedom on other lattice sites.
In presence of electron-electron interactions, the resulting local
effective action has an arbitrarily complicated form. Within DMFT,
the situation simplifies, and all the information about the environment
is contained in the local single particle spectral function $\Delta_{i}(\omega)$.
The calculation then reduces to solving an appropriate quantum impurity
problem supplemented by an additional self-consistency condition that
determines this {}``cavity function'' $\Delta_{i}(\omega)$.

The precise form of the DMFT equations depends on the particular model
of interacting electrons and/or the form of disorder, but most applications
\cite{georgesrmp} to this date have focused on Hubbard and Anderson
lattice models. The approach has been very successful in examining
the vicinity of the Mott transition in clean systems, in which it
has met spectacular successes in elucidating various properties of
several transition metal oxides \cite{vladgabisdmft2}, heavy fermion
systems, and even Kondo insulators \cite{re:Rozenberg96}.

\subsection{DMFT as an order-parameter theory for the MIT}

The central quantity in the DMFT approach is the local {}``cavity''
spectral function $\Delta_{i}(\omega)$. From the physical point of
view, this object essentially represents the {\em available electronic
states} to which an electron can {}``jump'' on its way out of a
given lattice site. As such, it provides a natural order parameter
description for the MIT. Of course, its form can be substantially
modified by either the electron-electron interactions or disorder,
reflecting the corresponding modifications of the electron dynamics.
According to Fermi's golden rule, the transition rate to a neighboring
site is proportional to the density of final states - leading to insulating
behavior whenever $\Delta_{i}(\omega)$ has a gap at the Fermi energy.
In the case of a Mott transition in the absence of disorder, such
a gap is a direct consequence of the strong on-site Coulomb repulsion,
and is the same for every lattice site.%
\begin{figure}[h]
 \centerline{\includegraphics[width=15cm]{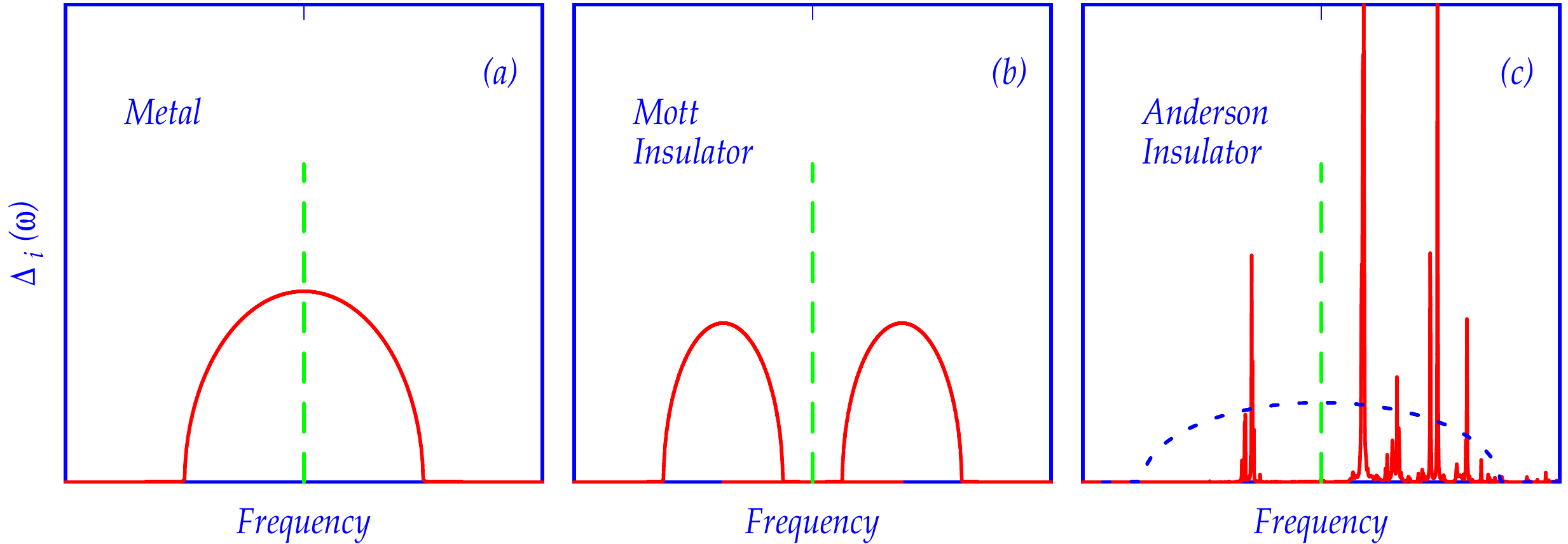}}

\caption{The local cavity spectral function $\Delta_{i}(\omega)$ as the order
parameter for the MIT. In a metal (a) there are available electronic
states near the Fermi level (dashed line) to which an electron from
a given site can delocalize. Both for a Mott insulator (b) and the
Anderson insulator (c) the Fermi level is in the gap, and the electron
cannot leave the site. Note that the {\em averaged} spectral function
(dotted line in (c)) has no gap for the Anderson insulator, and thus
cannot serve as an order parameter.}

\end{figure}

The situation is more subtle in the case of disorder-induced localization,
as first noted in the pioneering work of Anderson \cite{anderson58}.
Here, the \emph{average} value of $\Delta_{i}(\omega)$ has no gap
and thus cannot serve as an order parameter. However, as Anderson
noted a long time ago, {}``...no real atom is an average atom...''
\cite{andersonlocrev}. Indeed, in an Anderson insulator, the environment
{}``seen'' by an electron on a given site can be very different
from its average value. In this case, the \emph{typical} {}``cavity''
spectral function $\Delta_{i}(\omega)$ consists of several delta-function
(sharp) peaks, reflecting the existence of localized (bound) electronic
states, as shown in Fig. 1.4(c). Thus a \emph{typical} site is embedded
in an environment that has a \emph{gap} at the Fermi energy - resulting
in insulating behavior. We emphasize that the location and width of
these gaps strongly vary from site to site. These strong fluctuations
of the local spectral functions persist on the metallic side of the
transition, where the typical spectral density $\Delta_{typ}=\exp<\ln(\Delta_{i})>$
can be much smaller than its average value. Clearly, a full \emph{distribution
function} is needed to characterize the system. The situation is similar
as in other disordered systems, such as spin glasses \cite{re:Mezard86}.
Instead of simple averages, here the entire distribution function
plays a role of an order parameter, and undergoes a qualitative change
at the phase transition.

The DMFT formulation thus naturally introduces self-consistently defined
order parameters that can be utilized to characterize the qualitative
differences between various phases. In contrast to clean systems,
these order parameters have a character of distribution functions,
which change their qualitative form as we go from the normal metal
to the non-Fermi liquid metal, to the insulator.

\pagebreak

\section{Typical Medium Theory for Anderson localization}

\label{ch:anderson_localization}

In the following, we demonstrate how an appropriate local order parameter
can be defined and self-consistently calculated, producing a mean-field
like description of Anderson localization. This formulation is \emph{not
restricted} to either low temperatures or to Fermi liquid regimes,
and in addition can be straightforwardly combined with well-known
dynamical mean-field theories (DMFT) \cite{georgesrmp,dk-prl93,dk-prb94,motand,london,miranda-dobrosavljevic-rpp05}
of strong correlation. In this way, our approach which we call the
\emph{Typical Medium Theory} (TMT), opens an avenue for addressing
questions difficult to tackle by any alternative formulation, but
which are of crucial importance for many physical systems of current
interest.

Our starting point is motivated by the original formulation of Anderson
\cite{anderson58}, which adopts a \emph{local} point of view, and
investigates the possibility for an electron to \emph{delocalize}
from a given site at large disorder. This is most easily accomplished
by concentrating on the (unaveraged) local density of electronic states
(LDOS) \begin{equation}
\rho_{i}(\omega)=\sum_{n}\delta(\omega-\omega_{n})|\psi_{n}(i)|^{2}.\end{equation}
 In contrast to the global (averaged) density of states (ADOS) which
is not critical at the Anderson transition, the LDOS undergoes a qualitative
change upon localization, as first noted by Anderson \cite{anderson58}.
This follows from the fact that LDOS directly measures the local amplitude
of the electronic wavefunction. As the electrons localize, the local
spectrum turns from a continuous to an essentially discrete one \cite{anderson58},
but the \emph{typical} value of the LDOS vanishes. Just on the metallic
side, but very close to the transition, these delta-function peaks
turn into long-lived resonance states and thus acquire a finite \emph{escape
rate} from a given site. According to to Fermi's golden rule, this
escape rate can be estimated \cite{anderson58} as $\tau_{esc}^{-1}\sim t^{2}\rho$,
where $t$ is the inter-site hopping element, and $\rho$ is the density
of local states of the immediate neighborhood of a given site.

The \emph{typical} escape rate is thus determined by the typical local
density of states (TDOS), so that the TDOS directly determines the
conductivity of the electrons. This simple argument strongly suggests
that the TDOS should be recognized as an appropriate order parameter
at the Anderson transition. Because the relevant distribution function
for the LDOS becomes increasingly broad as the transition is approached,
the desired typical value is well represented by the \emph{geometric
average} $\rho_{{\rm TYP}}=\exp\{<\ln\rho>\}$. Interestingly, recent
scaling analyses \cite{re:Janssen98,re:Mirlin00} of the multi-fractal
behavior of electronic wavefunctions near the Anderson transition
has independently arrived at the same conclusion, identifying the
TDOS as defined by the geometric average as the fundamental order
parameter.

\subsection{Self-consistency conditions}

To formulate a self-consistent theory for our order parameter, we
follow the {}``cavity method,'' a general strategy that we borrow
from the DMFT \cite{georgesrmp}. In this approach, a given site is
viewed as being embedded in an effective medium characterized by a
local self energy function $\Sigma(\omega)$. For simplicity, we concentrate
on a single band tight binding model of noninteracting electrons with
random site energies $\varepsilon_{i}$ with a given distribution
$P(\varepsilon_{i})$. The corresponding local Green's function then
takes the form \begin{equation}
G(\omega,\varepsilon_{i})=[\omega-\varepsilon_{i}-\Delta(\omega)]^{-1}.\label{eq:ch4_2}\end{equation}
 Here, the {}``cavity function'' is given by \begin{equation}
\Delta(\omega)=\Delta_{o}(\omega-\Sigma(\omega))\equiv\Delta'+i\Delta'',\label{eq:ch4_3}\end{equation}
 and \begin{equation}
\Delta_{o}(\omega)=\omega-1/G_{o}(\omega),\label{eq:ch4_4}\end{equation}
 where the lattice Green's function \begin{equation}
G_{o}(\omega)=\int_{-\infty}^{+\infty}d\omega'\;\frac{\rho_{0}(\omega')}{\omega-\omega'}\label{eq:ch4_5}\end{equation}
 is the Hilbert transform of the bare density of states $\rho_{0}(\omega)$
which specifies the band structure.

Given the effective medium specified by a self-energy $\Sigma(\omega)$,
we are now in the position to evaluate the order parameter, which
we choose to be the TDOS as given by \begin{equation}
\rho_{{\rm typ}}(\omega)=\exp\left\{ \int d\varepsilon_{i}\; P(\varepsilon_{i})\;\ln\rho(\omega,\varepsilon_{i})\right\} ,\label{eq:ch4_6}\end{equation}
 where the LDOS $\rho(\omega,\varepsilon_{i})=-\frac{1}{\pi}{\rm Im}G(\omega,\varepsilon_{i})$,
as given by Eqs. \ref{eq:ch4_2}-\ref{eq:ch4_5}. To obey causality,
the Green's function corresponding to $\rho_{{\rm typ}}(\omega)$
must be specified by analytical continuation, which is performed by
the Hilbert transform \begin{equation}
G_{{\rm typ}}(\omega)=\int_{-\infty}^{+\infty}d\omega'\;\frac{\rho_{{\rm typ}}(\omega')}{\omega-\omega'}.\label{eq:ch4_7}\end{equation}

Finally, we close the self-consistency loop by setting the Green's
functions of the effective medium be equal to that corresponding to
the local order parameter, so that \begin{equation}
G_{{\rm em}}(\omega)=G_{o}(\omega-\Sigma(\omega))=G_{{\rm typ}}(\omega).\label{eq:ch4_8}\end{equation}

It is important to emphasize that our procedure defined by Eqs. \ref{eq:ch4_2}-\ref{eq:ch4_8}
is not specific to the problem at hand. The same strategy can be used
in any theory characterized by a local self-energy. The only requirement
specific to our problem is the definition of the TDOS as a local order
parameter given by Eq.\ref{eq:ch4_6} . If we choose the \emph{algebraic}
instead of the geometric average of the LDOS, our theory would reduce
to the well-known coherent potential approximation (CPA) \cite{re:Elliot74},
which produces excellent results for the ADOS for any value of disorder,
but finds no Anderson transition. Thus TMT is a theory having a character
very similar to CPA, with a small but crucial difference - the choice
of the correct order parameter for Anderson localization.%

\begin{figure}[h]
 \centerline{\includegraphics[width=10cm]{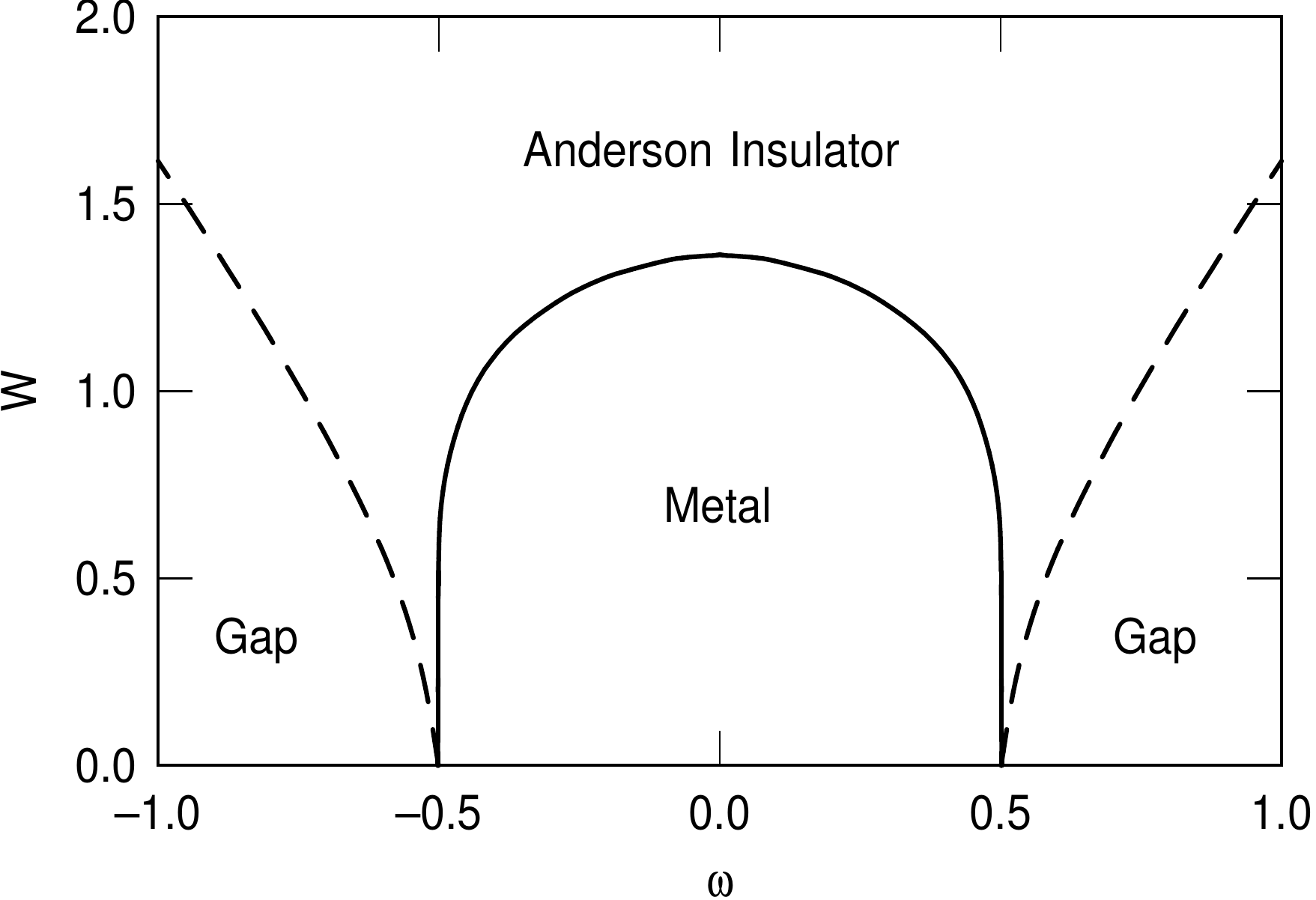}}

\caption{Phase diagram for the {}``semicircular'' model. The trajectories
of the mobility edge (full line) and the CPA band edge (dashed line)
are shown as a function the disorder strength $W$.}

\label{fig:ch4_2} 
\end{figure}

In our formulation, as in DMFT, all the information about the electronic
band structure is contained in the choice of the bare DOS $\rho_{0}(\omega)$.
It is not difficult to solve Eqs. \ref{eq:ch4_2}-\ref{eq:ch4_8}
numerically, which can be efficiently done using Fast Fourier transform
methods \cite{georgesrmp}. We have done so for several model of bare
densities of states, and find that most of our qualitative conclusions
do not depend on the specific choice of band structure. We illustrate
these findings using a simple {}``semicircular'' model for the bare
DOS given by $\rho_{0}(\omega)=\frac{4}{\pi}\sqrt{1-(2\omega)^{2}}$,
for which $\Delta_{o}(\omega)=G_{o}(\omega)/16$ \cite{georgesrmp}.
Here and in the rest of this paper all the energies are expressed
in units of the bandwidth, and the random site energies $\varepsilon_{i}$
are uniformly distributed over the interval $[-W/2,W/2]$. The evolution
of the TDOS as a function of $W$ is shown in Fig. \ref{fig:ch4_1}.
The TDOS is found to decrease and eventually vanish even at the band
center at $W\approx1.36$. For $W<W_{c}$, the part of the spectrum
where TDOS remains finite corresponds to the region of extended states
(mobile electrons), and is found to shrink with disorder, indicating
that the band tails begin to localize. The resulting phase diagram
is presented in Fig. \ref{fig:ch4_2}, showing the trajectories of
the mobility edge (as given by the frequency where the TDOS vanishes
for a given $W$, and the band edge where the ADOS as calculated by
CPA vanishes.

\subsection{Critical behavior}

Further insight in the critical behavior is obtained by noting that
near $W=W_{c}$ it proves possible to analytically solve Eqs. \ref{eq:ch4_2}-\ref{eq:ch4_8}.
Here we discuss the the critical exponent of the Anderson metal-insulator
transition within the TMT model. We will demonstrate that the critical
exponent $\beta$ with which the order parameter $\Delta''$ vanishes
at the transition is, in contradiction to the general expectations
\cite{goldenfeldbook}, non-universal in this model.

\subsubsection{Critical behavior in the middle of the band $\omega=0$}

\label{sec:ch4_middle_band} To start with, let us concentrate at
the band center ($\omega=0$), and expand Eqs. \ref{eq:ch4_2}-\ref{eq:ch4_8}
in powers of the order parameter $\Delta''$. In the limit of $\omega=0$
self-consistency equations quantities $\Delta,\overline{G}$, and
$\Sigma$ become purely imaginary, and near the critical disorder
typical Green's function can be expanded in powers of the parameter
$\Delta''$: 

\begin{eqnarray}
\overline{G(\omega,\varepsilon_{i})} & = & i\Delta''=\left<\frac{\Delta''}{(\omega-\varepsilon_{i}-\Delta')^{2}+\Delta''^{2}}\right>_{{\rm typ}}\\
 & = & i\Delta''\exp\left[-\int d\varepsilon P(\varepsilon_{i})\log[\varepsilon_{i}^{2}+\Delta''^{2}]\right]=i\Delta''f(\Delta'')\approx i\Delta''(a-b\Delta'')\nonumber \end{eqnarray}
 where \begin{eqnarray}
a & = & f(0)=\exp\left[-2\int d\varepsilon P(\varepsilon)\log\left|\varepsilon\right|\right]\\
b & = & \frac{\partial f}{\partial\Delta}\Biggr|_{\Delta=0}=a\cdot\exp\left[-2\int d\varepsilon P(\varepsilon)\frac{-2\Delta''}{\varepsilon^{2}+\Delta''^{2}}\right]\nonumber \\
 & = & -a\int d\varepsilon P(\varepsilon)2\pi\delta(\varepsilon)=-2\pi aP(0),\end{eqnarray}
 and after trivial algebraic operations our self-consistency equations
\ref{eq:ch4_2}-\ref{eq:ch4_8} reduce to a single equation for the
order parameter $\Delta''$ \begin{eqnarray}
\Delta''=\frac{\Delta''}{t^{2}}(a-b\Delta'')\int_{-2t}^{2t}\rho_{0}(\varepsilon)\varepsilon^{2}d\varepsilon.\label{eq:ch4_12}\end{eqnarray}
 Equation \ref{eq:ch4_12} shows that near the transition along $\o=0$
direction our order parameter $\Delta''$ vanishes linearly (critical
exponent $\beta=1$) independently of the choice of bare lattice DOS
$\rho_{0}$. In specific case of semicircular bare DOS, where \begin{eqnarray}
\Delta''=a\Delta''-b\Delta''^{2}.\end{eqnarray}
 the transition where $\Delta''$ vanishes is found at $a=1$, giving
$W=W_{c}=e/2=1.3591$, consistent with our numerical solution. Near
the transition, to leading order \begin{equation}
\rho_{typ}(W)=-\frac{\Delta''}{\pi}=\left(\frac{4}{\pi}\right)^{2}(W_{c}-W),\end{equation}

\subsubsection{Critical behavior near the band edge $\omega=\omega_{c}$}

In order to analytically examine scaling the critical behavior at
finite $\omega$, for simplicity we focus on a semi-circular bare
DOS, where self-consistency Eqs \ref{eq:ch4_2}-\ref{eq:ch4_8} are
greatly simplified \begin{eqnarray}
\overline{G(\omega,\varepsilon_{i})} & = & \Delta'+i\Delta''\quad\Rightarrow\quad\Delta''=-\pi\rho_{{\rm typ}}=Im\overline{G(\omega,\varepsilon_{i})}\\
\Delta''(\omega) & = & -\exp\left\{ \int d\varepsilon_{i}P(\varepsilon_{i})\ln\left[\frac{-\Delta''}{(\omega-\varepsilon_{i}-\Delta')+\Delta''^{2}}\right]\right\} \label{eq:ander_crit1}\\
\Delta'(\omega) & = & -H[\Delta''(\omega)],\end{eqnarray}
 however as in previous section, we expect the critical exponent to
be the same for \textit{any} bare DOS.

To find the general critical behavior near the mobility edge, we need
to expand Eq. \ref{eq:ander_crit1} in powers of $\Delta''$ \begin{eqnarray}
\Delta''=\Delta''\exp\left\{ -\int d\varepsilon_{i}P(\varepsilon_{i})\ln\left[(\omega-\varepsilon_{i}-\Delta')^{2}+\Delta''^{2}\right]\right\} \equiv\Delta''f(\Delta''),\label{eq:and_crit3}\end{eqnarray}
 which cannot be done explicitly, since $\Delta'$ and $\Delta''$
are related via Hilbert transform, which depends on the entire function
$\Delta''(\omega)$, and not only on its form near $\omega=\omega_{c}$.
Nevertheless the quantity $\omega-\Delta'(\omega)$ assumes a well-defined
$W$-dependent value at the mobility edge $\omega_{c}=\omega_{c}(W)$,
making it possible for us to determine a \emph{range} of values a
critical exponent $\beta$ may take.

After expanding $f(\Delta'')=1$ defined by Eq. \ref{eq:and_crit3}
\begin{eqnarray}
f(\Delta'') & = & a-b\Delta''+O\left(\Delta''^{2}\right)\nonumber \\
a & = & f(0)=\exp\left\{ -2\int d\varepsilon P(\varepsilon)\ln|\omega-\varepsilon-\Delta'(\omega)|\right\} \nonumber \\
b & = & \frac{\partial f}{\partial\Delta''}\Biggr|_{\Delta''=0}=a\lim_{\Delta''\rightarrow0}\left[\int d\varepsilon P(\varepsilon)\frac{2\Delta''}{(\omega-\varepsilon-\Delta')^{2}+\Delta''^{2}}\right]\nonumber \\
 & = & a\int d\varepsilon P(\varepsilon)2\pi\delta(\omega-\varepsilon-\Delta')=2\pi aP(\omega-\Delta'),\end{eqnarray}
 we find that to the leading order $\Delta''$ has the following $\omega$
dependence ($\delta a\equiv1-a$) \begin{eqnarray}
\Delta'' & = & \frac{1}{2\pi P(\omega-\Delta')}\left[\frac{1}{a}-1\right]\nonumber \\
 & \approx & \frac{1}{2\pi P(\omega_{c}-\Delta'(\omega_{c}))}\delta a(\omega)\propto\delta\omega^{\beta}.\label{eq:ander_crit_d1}\end{eqnarray}

The functional form of $\delta a(\omega)$ is readily found \begin{eqnarray}
a & = & \exp\left\{ -2\int d\varepsilon P(\varepsilon)\ln|\omega-\varepsilon-\Delta'(\omega)|\right\} \\
\delta a(\omega) & = & 2\int d\varepsilon P(\varepsilon)\frac{1}{\omega-\varepsilon-\Delta'(\omega)}(\delta\omega-\delta\Delta'(\omega)),\label{eq:ander_crit_d2}\end{eqnarray}
 and combining Eqs. \ref{eq:ander_crit_d1},\ref{eq:ander_crit_d2}
we arrive to \begin{eqnarray}
\Delta''=\Delta''_{0}(\delta\omega-\delta\Delta')\qquad\Delta''_{0}=\frac{1}{\pi P(\omega_{c}-\Delta'(\omega_{c}))}\int d\varepsilon\frac{P(\varepsilon)}{\omega_{c}-\varepsilon-\Delta'(\omega_{c})}.\label{eq:ander_crit_d3}\end{eqnarray}
 Note that $\delta\omega$ is negative, since in the range of interest
$\omega <\omega_{c}$. In Eq \ref{eq:ander_crit_d3} $\int d\varepsilon\frac{P(\varepsilon)}{\omega_{c}-\varepsilon-\Delta'(\omega_{c})}$
is the Hilbert transform of $P(\varepsilon)$, which is positive for
$\omega_{c}-\Delta'>0$ (right band edge), and it is negative for
the left one, where $\delta\omega >0$.

\begin{figure}[h]
\centerline{\includegraphics[width=4in]{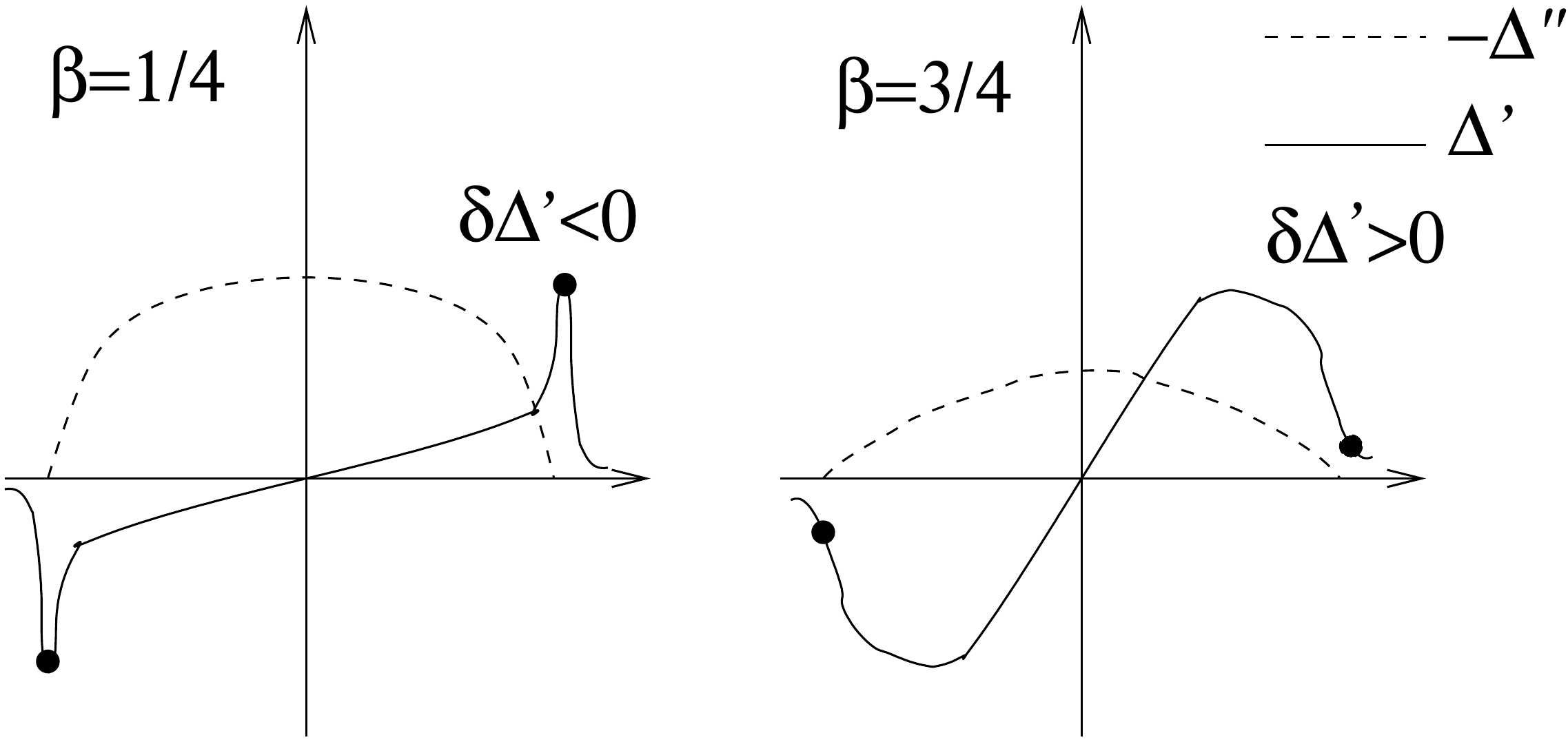}}

\caption{Requirement on $\delta\Delta'$ to be positive definite forces the
value of critical exponent $\beta$ to be larger than $1/2$}.

\label{fig:critical_exponent} 
\end{figure}

The lower bound on critical exponent $\beta$ is 0, to insure that
$\Delta''$ is convergent and vanishing at $\omega=\omega_{c}$. Now,
if we were to assume that the leading contribution to $\Delta''$
comes from $\delta\omega$ (and $\delta\Delta'$ can be neglected),
the conclusion would be that $\Delta''\propto\delta\omega$, and the
critical exponent $\beta=1$, just like in $\omega =0$ case. However,
this value of $\beta$ is unphysical, since the Kramers-Kroning predicts
$\Delta'$ to be logarithmically divergent ($\Delta'\gg\delta\omega$)
when $\Delta''\propto\delta\omega$. This is in direct contradiction
with our initial statement of $\delta\omega$ being a leading contribution
in Eq. \ref{eq:ander_crit_d3} ($|\delta\Delta'|\ll|\delta\o|$),
and we conclude that $\delta\Delta'\propto\delta\omega^{\beta}$ is
the leading contribution \begin{eqnarray}
\Delta''\approx-\Delta''_{0}\delta\Delta'\propto-\delta\omega^{\beta}\end{eqnarray}
 with $\beta\in(0,1)$. This $\Delta''$ being a negative definite
quantity imposes a constraint $\delta\Delta'>0$, which is only satisfied
for $\beta >1/2$ (see Fig. \ref{fig:critical_exponent}, thus our critical
exponent can vary in the range $\beta\in(1/2,1)$.

Although general arguments for second-order phase transitions \cite{goldenfeldbook}
predict universality of exponent $\beta$, we find the exponent is
non-universal, which is not uncommon in some special cases of mean
field theories\cite{re:Sompolinsky82}. It plausible that this critical
exponent anomaly can be remedied if the MFT is extended to incorporate
long range fluctuations effects beyond mean-field theory, but this
remains an open problem for future work.

\begin{figure}[h]

\centerline{\includegraphics[width=10cm]{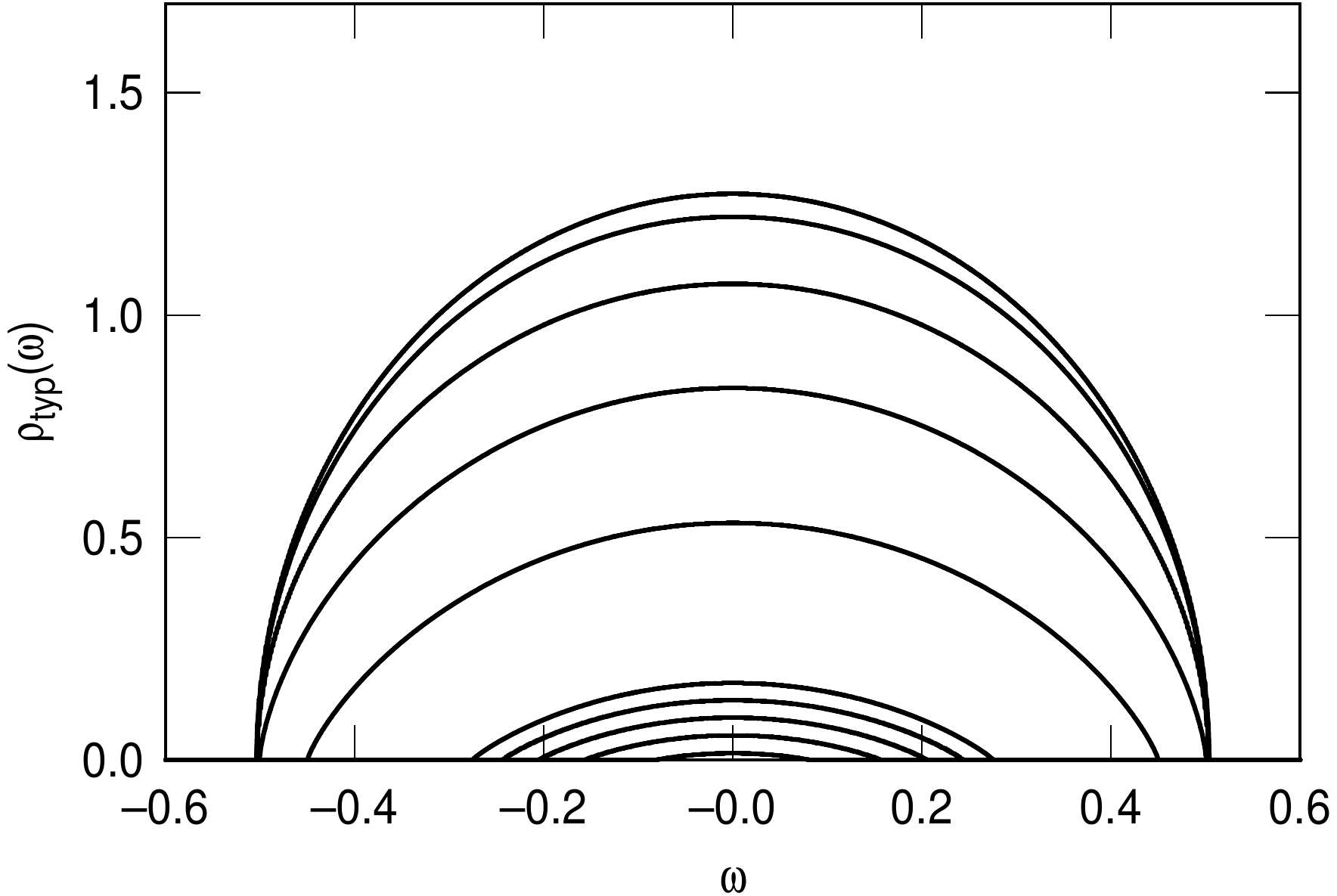}} 
\caption{Typical density of states for for the SC model, for disorder values
$W$ = 0, 0.25, 0.5, 0.75, 1, 1.25, 1.275, 1.3, 1.325, 1.35. The entire
band localizes at $W=W_{c}=e/2\approx1.359$.}

\label{fig:ch4_1} 
\end{figure}

\subsubsection{Scaling behavior near the critical disorder $W=W_{c}$.}

The complete analytical solution for TDOS is difficult to obtain for
arbitrary $\omega$ and $W$. Still, the approach discussed in section
\ref{sec:ch4_middle_band} can be extended to find a full frequency-dependent
solution $\rho_{{\rm typ}}(\omega,W)$ close to the critical value
of disorder $W=W_{c}$ and which assumes a simple scaling form \begin{equation}
\rho_{{\rm typ}}(\omega,W)=\rho_{o}(W)f\left(\omega/\omega_{o}(W)\right).\end{equation}

\begin{figure}[h]
 \centerline{\includegraphics[width=6.15cm]{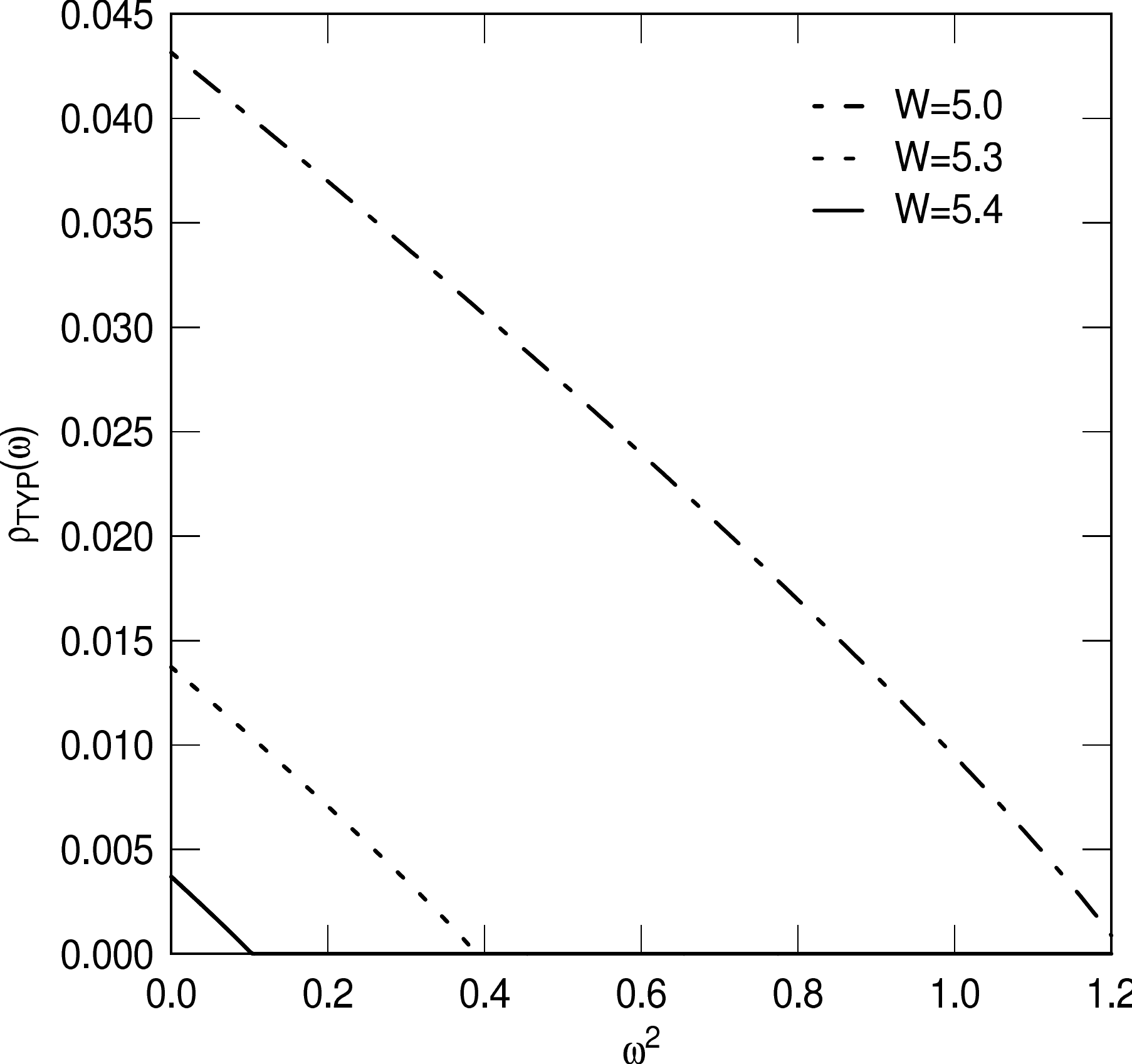}\includegraphics[width=6cm]{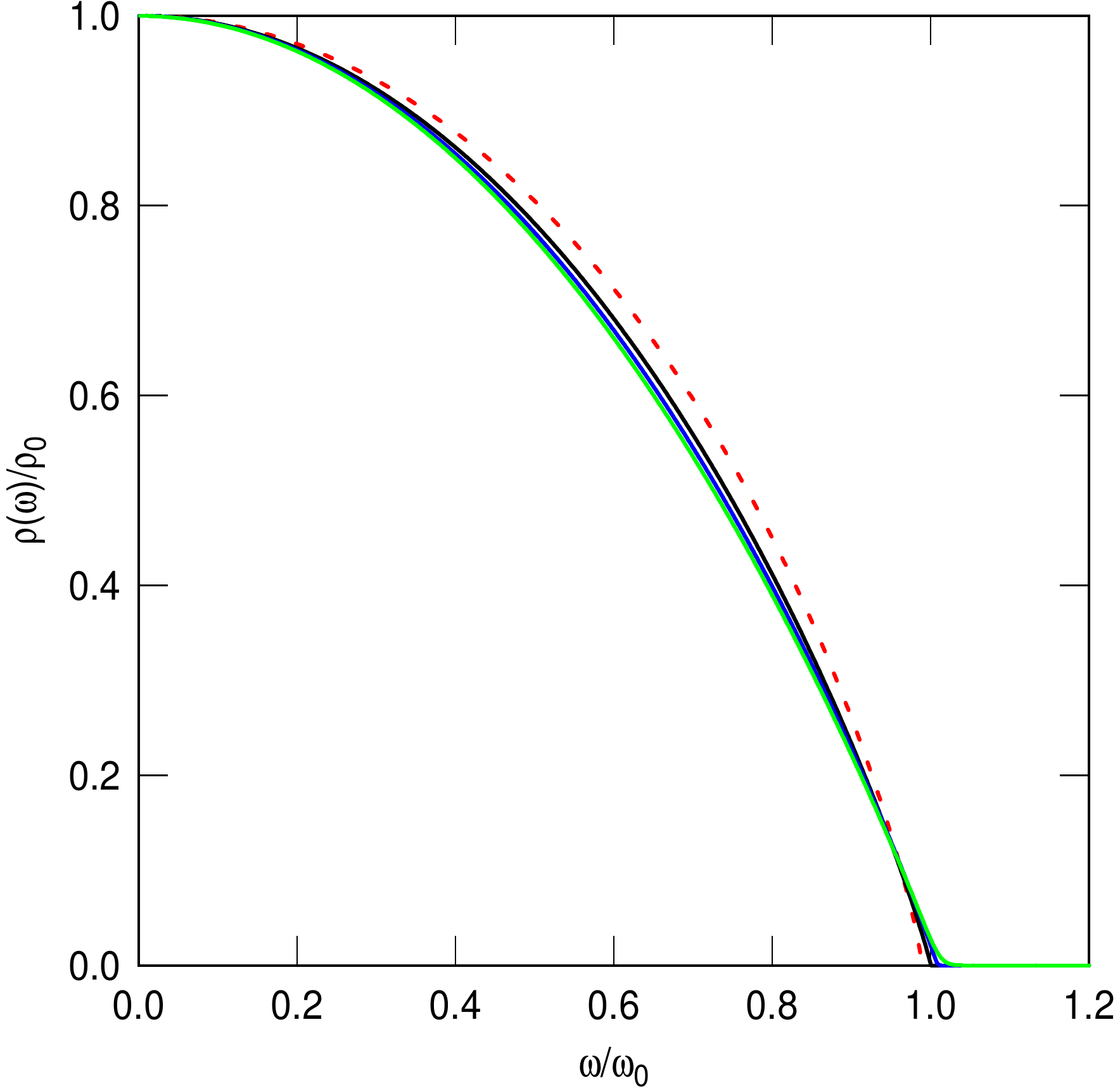}}

\caption{Left: unscaled typical DOS for various disorder displays parabolic
behavior near the MIT. Right: scaling behavior near the critical disorder.
The range of disorders where parabolic behavior is observed is, in
fact, quite broad - $W\in(1,W_{c})$, $W_{c}=e/2$.}

\label{fig:dos_scaled} 
\end{figure}

Our numerical solution (see Figs. \ref{fig:ch4_1}, \ref{fig:dos_scaled})
has suggested that the corresponding scaling function assumes a simple
{\em parabolic} form $f(x)=1-x^{2}$ 
\begin{eqnarray}
\rho_{{\rm tyo}}(\omega)\approx\rho_{0}\left(1-\frac{\omega^{2}}{\omega_{0}^{2}}\right)\qquad|\omega|<|\omega_{0}|.\label{eq:scal_ansatz}
\end{eqnarray}
\begin{figure}[h]
\centerline{\includegraphics[width=9cm]{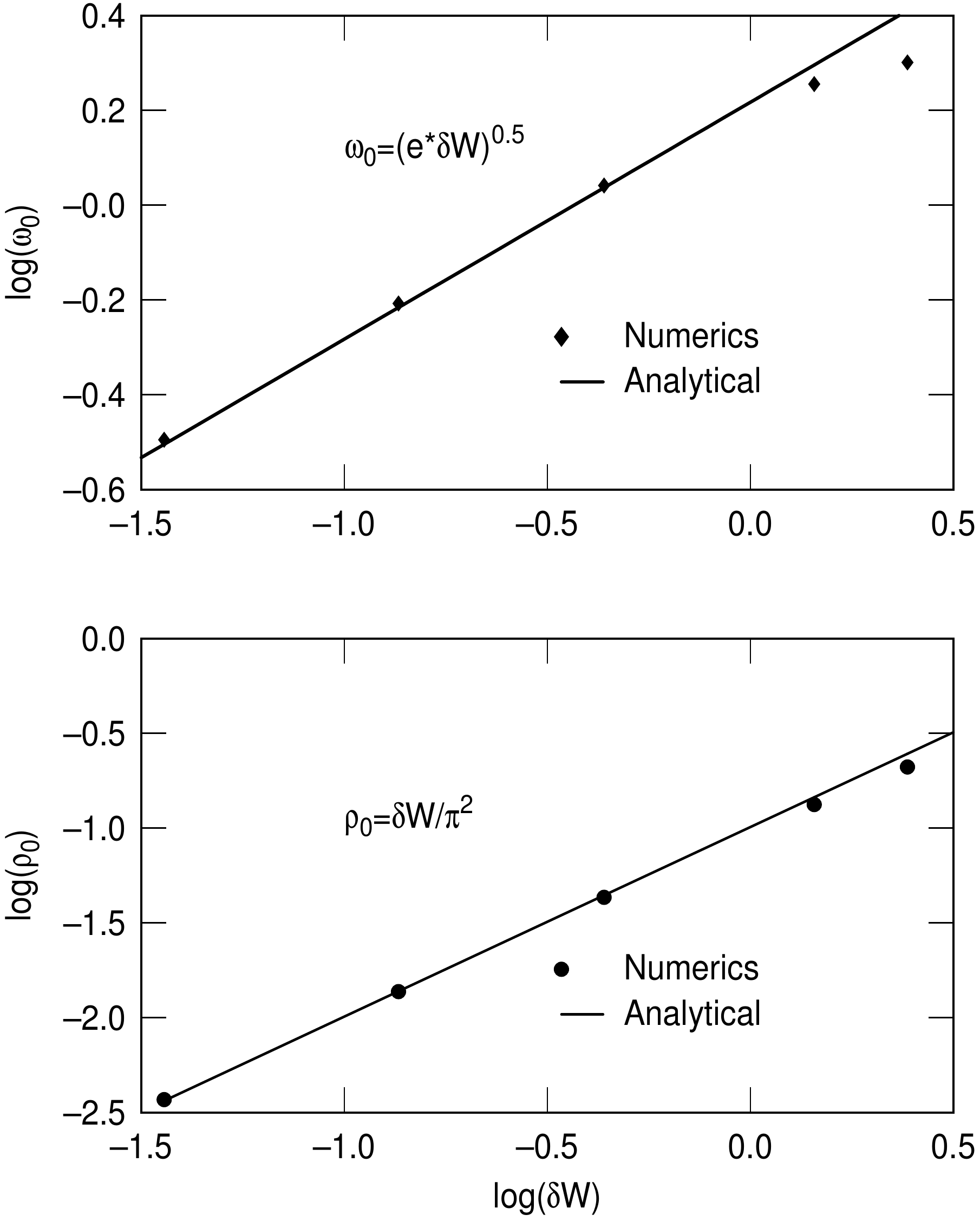}}

\caption{Scaling parameters $\rho_{0}$ and $\omega_{0}$ as functions of the
distance to the transition $\delta W=W_{c}-W$. Numerically obtained
values ($\diamond$ and $\bullet$) are in good agreement with analytical
predictions (full line).}

\label{fig:scal_parameters} 
\end{figure}
 In the following, we analytically calculate the scaling parameters
$\rho_{0}$ and $\omega_{0}$ for semicircular bared DOS and box distribution
of disorder \begin{equation}
P(\varepsilon)=\left\{ \begin{array}{ll}
\frac{1}{W}\qquad & \varepsilon\in[-\frac{W}{2},\frac{W}{2}]\\
0\qquad & \varepsilon\not\in[-\frac{W}{2},\frac{W}{2}].\end{array}\right.\end{equation}

\begin{eqnarray*}
\Delta'' & = & -\exp\left[\int d\varepsilon P(\varepsilon)\log\left(-\frac{\Delta''}{[(\omega-\varepsilon-\Delta')^{2}+\Delta''^{2}]}\right)\right]\end{eqnarray*}
 after averaging over disorder takes the following form \begin{eqnarray}
2W & = & a_{-}\log(\Delta''^{2}+a_{-}^{2})+a_{+}\log(\Delta''^{2}+a_{+}^{2})\nonumber \\
 &  & +2\Delta''\left[\arctan\left(\frac{a_{+}}{\Delta''}\right)+\arctan\left(\frac{a_{-}}{\Delta''}\right)\right],\label{eq:quad_fit}\end{eqnarray}
 where \begin{eqnarray}
a_{\pm}=\frac{W}{2}\pm(\Delta'-\omega).\end{eqnarray}

Exact expression for the real part of the cavity field $\Delta'$
is obtained by performing a Hilbert transformation of ansatz \ref{eq:scal_ansatz}:
\begin{eqnarray}
\Delta'' & = & -\pi\rho_{0}\left(1-\frac{\omega^{2}}{\omega_{0}^{2}}\right)\label{eq:quad_deltas}\\
\Delta' & = & -H[\Delta'']=\rho_{0}\left(2\frac{\omega_{0}\omega}{\omega_{0}^{2}}-(1-\frac{\omega^{2}}{\omega_{0}^{2}})\log\left|\frac{\omega-\omega_{0}}{\omega-\omega_{0}}\right|\right).\nonumber \end{eqnarray}

Expanding Eqs. \ref{eq:quad_fit},\ref{eq:quad_deltas} to the second
order in small $\omega$ results in a system of equations: \begin{eqnarray*}
 &  & \frac{2\pi\rho_{0}}{W}\arctan\left(\frac{W}{2\pi\rho_{0}}\right)+\frac{1}{2}\log\left(\frac{W^{2}}{4}+\pi^{2}\rho_{0}^{2}\right)=1\\
 &  & \frac{2\pi\rho_{0}}{W\omega_{0}^{2}}\arctan\left(\frac{W}{2\pi\rho_{0}}\right)=\frac{\left(\frac{4\rho_{0}}{\omega_{0}}-1\right)^{2}}{2\left(\frac{W^{2}}{4}+\pi^{2}\rho_{0}^{2}\right)},\end{eqnarray*}
 which can be solved for scaling parameters used the in original ansatz, Eq. \ref{eq:scal_ansatz}.
\begin{eqnarray}
\rho_{0} & = & \frac{4(W_{c}-W)}{\pi^{2}}\\
\omega_{0} & = & \sqrt{\frac{e}{2}}\sqrt{W_{c}-W}.\end{eqnarray}

\subsection{Numerical test of TMT}

\begin{figure}[h]
\centerline{\includegraphics[width=10cm]{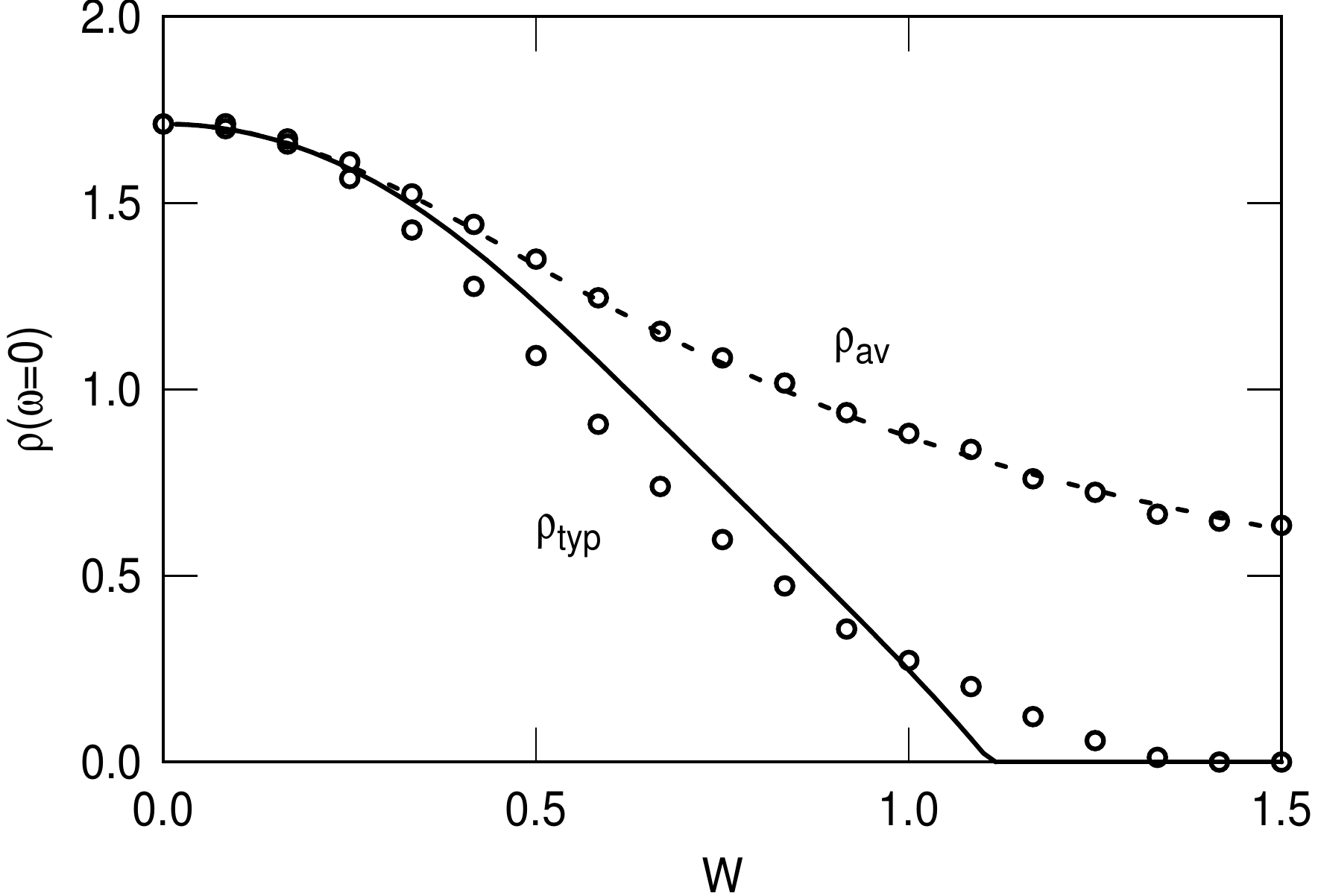}}

\caption{Typical and average DOS for a three dimensional cubic lattice at the
band center ($\omega=0$). Results from first-principle numerical
calculations (circles) are compared to the predictions of TMT (for
TDOS - full line ) and CPA (for ADOS - dashed line).}

\label{fig:ch4_3} 
\end{figure}

In order to gauge the quantitative accuracy of out theory, we have
carried out first-principles numerical calculations for a three dimensional
cubic lattice with random site energies, using exact Green functions
for an open finite sample attached to two semi-infinite clean leads
\cite{tmt}. We computed both the average and the typical DOS at the
band center as a function of disorder, for cubes of sizes $L$ =4,
6, 8, 10, 12, and 16, and averages over 1000 sample realizations,
in order to obtain reliable data by standard finite size scaling procedures.
The TMT and CPA equations for the same model were also solved by using
the appropriate bare DOS (as expressed in terms of elliptic integrals),
and the results are presented in Fig. \ref{fig:ch4_3}.

\pagebreak
We find remarkable agreement between these numerical data \cite{tmt}
and the self-consistent CPA calculations for the ADOS, but also a
surprisingly good agreement between the numerical data and the TMT
predictions for the TDOS order parameter. For a cubic lattice, the
exact value is $W_{c}\approx1.375$ \cite{re:Grussbach95}, whereas
TMT predicts a 20\% smaller value $W_{c}\approx1.1$. The most significant
discrepancies are found in the critical region, since TMT predicts
the order parameter exponent $\beta=1$, whereas the exact value is
believed to be $\beta\approx1.5$, consistent with our numerical data.
Argument based on the multi-fractal scaling analysis \cite{re:Janssen98,re:Mirlin00},
together with numerical calculations\cite{re:Grussbach95} of the
multi-fractal spectra of wavefunction amplitudes have suggested that
in three dimensions, the TDOS order parameter exponent $\beta$ should
be equal to the conductivity exponent $\mu\approx1.5$. The result
$\beta=\mu=1+O(\varepsilon)$ is also found within the $2+\epsilon$
approach \cite{re:Janssen98,re:Mirlin00}. Nevertheless, we conclude
that TMT is as accurate as one can expect from a simple mean-field
formulation. 
\subsection{Transport properties}

Most previous conventional transport theories, while providing a wonderful
description of good metals, fail to describe the transport properties
of highly disordered materials.

In most metals, the temperature coefficient of resistivity (TCR) $\alpha$
is positive, because phonon scattering decreases the electronic mean
free path as the temperature is raised. The sign of TCR \begin{eqnarray}
\alpha=\frac{d\ln\rho_{res}(T)}{dT}\end{eqnarray}
 can be deduced from Matthiessen's rule which asserts that the total
resistivity in the presence of two or more scattering mechanisms is
equal to the sum of the resistivities that would result if each mechanism
were the only one operating, for example: \begin{eqnarray}
\rho_{res}=\rho_{res}^{(1)}+\rho_{res}^{(2)}.\label{eq:matthiessen}\end{eqnarray}

Matthiessen's rule, as stated in Eq \ref{eq:matthiessen} follows
from the Boltzmann equation with the assumption of a wave-vector-independent
relaxation time for each scattering mechanism, so if $\rho_{0}$ is
the resistivity of a disordered metal at zero temperature and $\rho_{ph}(T)$
is the resistivity of the ordered material due to electron-phonon
scattering, then the total resistivity at finite temperature is \begin{eqnarray}
\rho(T)=\rho_{0}+\rho_{ph}(T)\geq\rho_{0}\end{eqnarray}
 predicting that the TCR is positive ($\alpha>0$)

Mooij \cite{re:Mooij73} in 1973 have pointed out that there exist
many highly disordered metals which are poor conductors and have $\rho(T)\leq\rho_{0}$
and negative TCRs, which clearly violate Matthiessen's rule. In fact
in these materials the Boltzmann-equation formalism itself is breaking
down. Apparently what is happening is that, because of the strong
disorder and resulting multiple correlated scattering, the Boltzmann
hypothesis of independent scattering events fails. The simple picture
of temperature fluctuations impeding transport of electrons (positive
TCR) is now replaced with the temperature fluctuations releasing the
localized electrons and increasing the conductivity (negative TCR).
When the transport properties are addressed within the TMT, the interplay
of several localization mechanisms is considered, which is capable
of producing the negative TCRs observed in highly disordered materials.

We start addressing the transport properties of our system within
the TMT by pointing out that the escape rate from a given site can
be rigorously defined in terms of the cavity field (see Eq. \ref{eq:ch4_2}),
and using our solution of the TMT equations, we find $\tau_{esc}^{-1}=-{\rm Im}\Delta(0)\sim\rho_{{\rm TYP}}\sim(W_{c}-W)$.
To calculate the conductivity within our local approach, we follow
a strategy introduced by Girvin and Jonson (GJ) \cite{re:Girvin80},
who pointed out that close to the localization transition, the conductivity
can be expressed as $\sigma=\Lambda a_{12}$, where $\Lambda$ is
a vertex correction that represents hops to site outside of the initial
pair $i$ and $j$, and $a_{12}$ is a two-site contribution to the
conductivity, that can be expressed as \begin{equation}
a_{12}=<A_{12}A_{21}-A_{11}A_{22}>,\label{eq:ch4_100}\end{equation}
 where $A_{ij}=-{\rm Im}G_{ij}$ is the spectral function corresponding
to the nearest neighbor two-site cluster, $<\cdots>$ represents the
average over disorder.%
\begin{figure}[h]
 \centerline{\includegraphics[width=11cm]{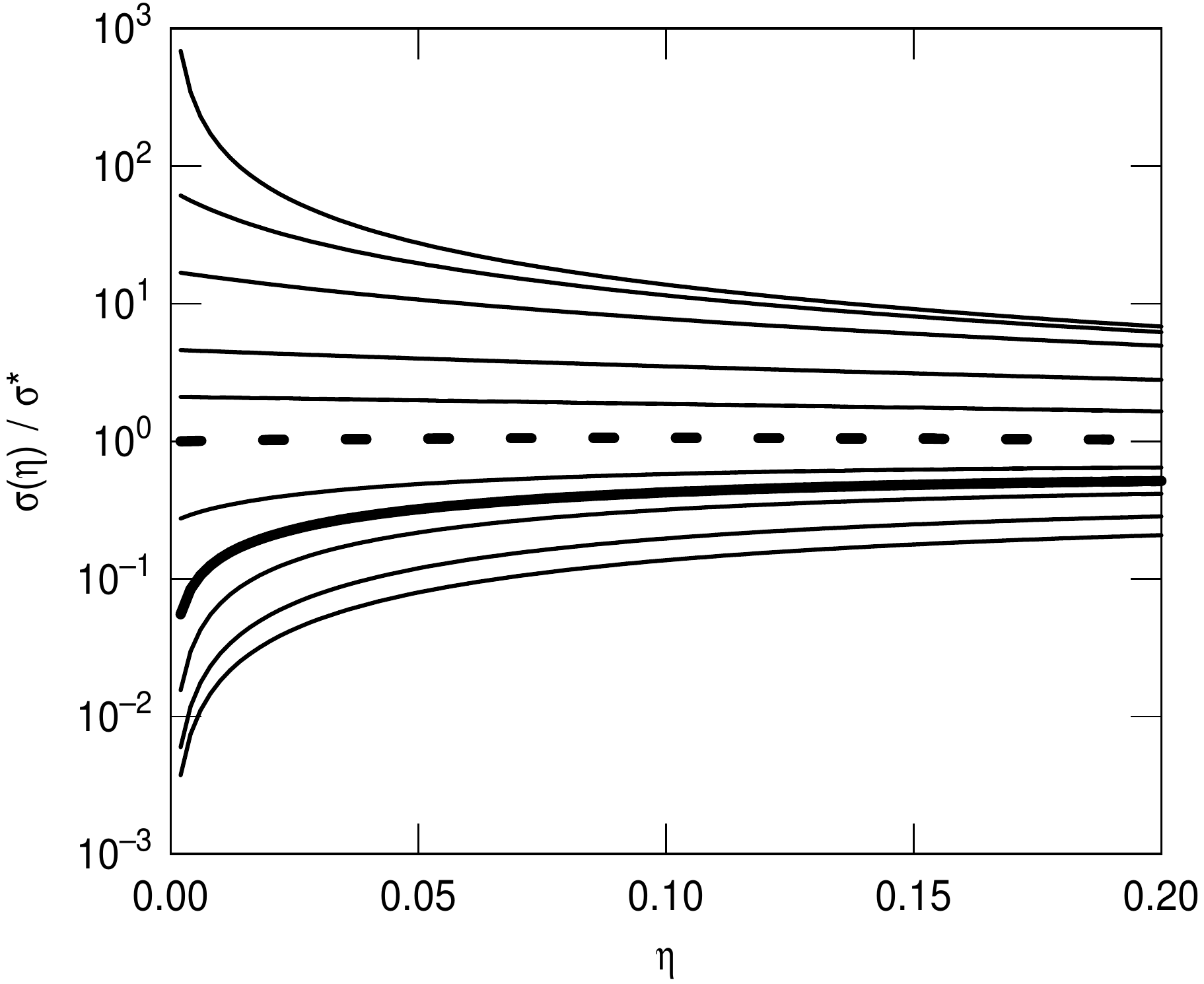}}

\caption{Conductivity as a function of the inelastic scattering rate $\eta$
for for the SC model at the band center and $W$ = 0, 0.125, 0.25,
0.5, 0.75, 1, 1.25, \textbf{1.36}, 1.5, 1.75, 2. The {}``separatrix''
($\sigma=\sigma^{*}$ independent of $\eta$, i. e. temperature) is
found at $W=W^{*}\approx1$ (dashed line). The critical conductivity
$\sigma_{c}(\eta)\sim\eta^{1/2}$ corresponds to $W=W_{c}=1.36$ (heavy
full line).\label{fig:ch4_4}}

\end{figure}

We examine the temperature dependence of the conductivity as a function
of $W$. Physically, the most important effect of finite temperatures
is to introduce finite inelastic scattering due to interaction effects.
At weak disorder, such inelastic scattering increases the resistance
at higher temperatures, but in the localized phase it produces the
opposite effect, since it suppresses interference processes and localization.
To mimic these inelastic effects within our noninteracting calculation,
we introduce by hand an additional scattering term in our self-energy,
viz. $\Sigma\rightarrow\Sigma-i\eta$ or it can be treated as the
imaginary part of $\omega\rightarrow\omega+\i\eta$. The parameter
$\eta$ measures the inelastic scattering rate, and is generally expected
to be a monotonically increasing function of temperature.

The relevant $\eta$-dependent Green's functions $G_{ij}$ \begin{eqnarray}
G_{ii} & = & \frac{\omega+i\eta-\varepsilon_{i}-\Delta}{(\omega+i\eta-\varepsilon_{i}-\Delta)(\omega+i\eta-\varepsilon_{j}-t^{2})-t^{2}}\nonumber \\
G_{ij} & = & \frac{t}{(\omega+i\eta-\varepsilon_{i}-\Delta)(\omega+i\eta-\varepsilon_{j}-t^{2})-t^{2}},\end{eqnarray}
 reduce expression \ref{eq:ch4_100} to an integrable form \begin{eqnarray}
 &  & a_{12}=4\frac{(\Delta''-\eta)^{2}}{W^{2}}\times\nonumber \\
 &  & \int_{\omega-\Delta'-W/2}^{\omega-\Delta'+W/2}{\frac{dx}{x^{2}+(\Delta''-\eta)^{2}}\arctan\left[\frac{-x+y(x^{2}+(\Delta''-\eta)^{2})}{(\Delta''-\eta)(x^{2}+(\Delta''-\eta)^{2}+1)}\right]\Bigg|_{y=\omega-\Delta'-W/2}^{y=\omega-\Delta'+W/2}}\end{eqnarray}
 that can be solved numerically as a function of temperature $\eta$
and disorder $W$.

We have computed $a_{12}$ by examining two sites embedded in the
effective medium defined by TMT ($\Delta_{{\rm TMT}}$), thus allowing
for localization effects. The vertex function $\Lambda$ remains \emph{finite}
at the localization transition \cite{girvinjonson}, and thus can
be computed within. We have used the CPA approach to evaluate the
vertex function as $\Lambda=\sigma_{cpa}/a_{12}^{cpa}$, where $\sigma_{cpa}$
is the CPA conductivity calculated using approach described by Elliot
\cite{re:Elliot74} \begin{eqnarray}
\sigma(\omega)\propto\int_{-B/2}^{B/2}{d\varepsilon\rho_{0}(\varepsilon)Im[G(\omega,\varepsilon)]^{2}}\\
\rho_{0}(\varepsilon)=\left(\frac{B^{2}}{4}-\varepsilon^{2}\right)^{3/2},\end{eqnarray}
 and $a_{12}^{cpa}$ is the two-site correlation function embedded
in the CPA effective medium($\Delta_{{\rm CPA}}$). Since TMT reduces
to CPA for weak disorder, our results reduce to the correct value
at $W\ll W_{c}$, where the conductivity reduces to the Drude-Boltzmann
form. The resulting critical behavior of the $T=0$ conductivity follows
that of the order parameter, $\sigma\sim\rho_{{\rm TYP}}\sim(W_{c}-W)$,
giving the conductivity exponent $\mu$ equal to the order parameter
exponent $\beta$, consistent with what is expected.

The resulting dependence of the conductivity as a function of $\eta$
and $W$ is presented in Fig. \ref{fig:ch4_4}. As $\eta$ (i. e.
temperature) is reduced, we find that the conductivity curves {}``fan
out'', as seen in many experiment close to the MIT \cite{leeramakrishnan,abrahams-rmp01}.
Note the emergence of a {}``separatrix''\cite{leeramakrishnan,abrahams-rmp01}
where the conductivity is temperature independent, which is found
for $W\approx1$, corresponding to $k_{F}\ell\sim2$, consistent with
some experiments \cite{leeramakrishnan}. At the MIT, $\sigma_{c}(\eta)\sim\rho_{{\rm TYP}}(\eta)\sim\eta^{1/2}$.

\section{Mott-Anderson Transitions}

\subsection{Two-fluid picture of Mott}

A first glimpse of the basic effect of disorder on the Mott transition
at half filling was outlined already by Mott,~\cite{mott-book90}
who pointed out that important differences can be seen even in the
strongly localized (atomic) limit.%
\begin{figure}[h]

\begin{centering}
\includegraphics[width=12cm]{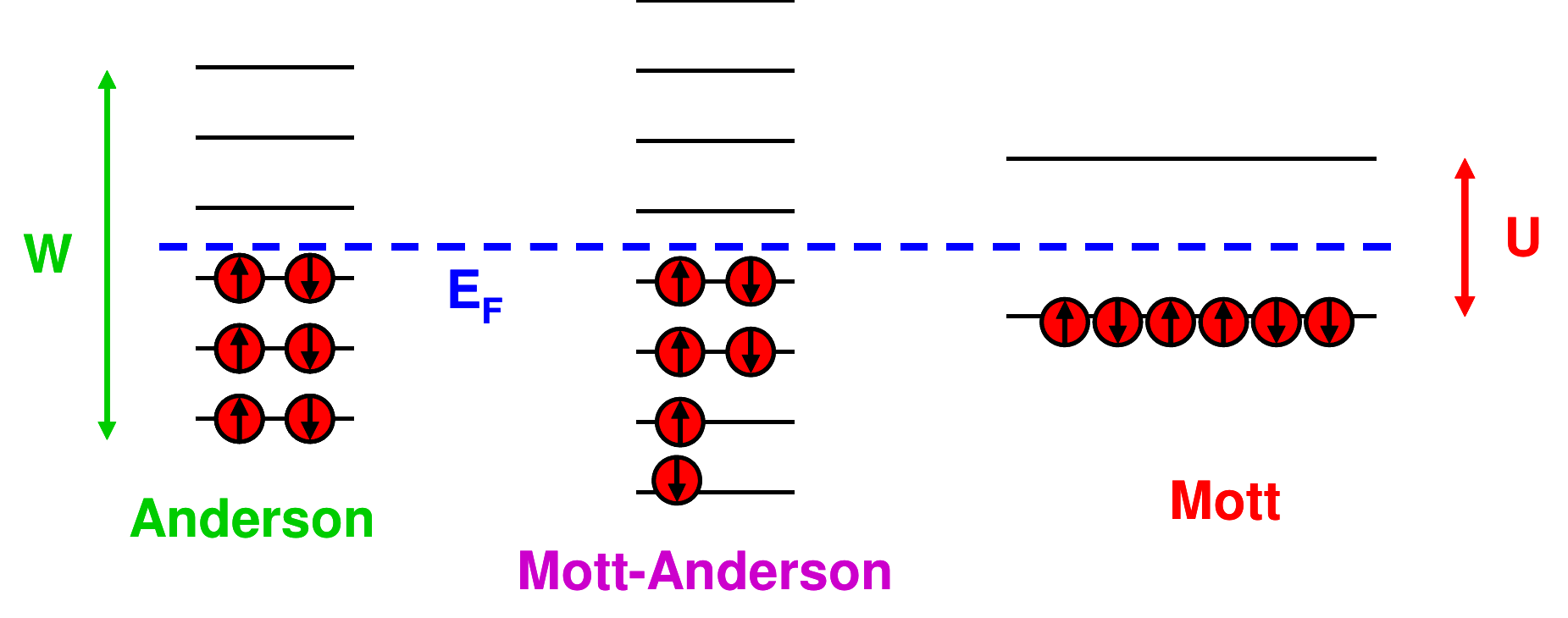} 
\par\end{centering}

\caption{Energy level occupation in the strongly localized (atomic) limit.
In a Mott-Anderson insulator (center), the disorder strength $W$
is larger then the Coulomb repulsion $U$, and a two-fluid behavior
emerges. Here, a fraction of localized states are doubly occupied
or empty as in an Anderson insulator. Coexisting with those, other
states remain singly occupied forming local magnetic moments, as in
a Mott insulator. }

\end{figure}

For weak to moderate disorder $W<U$, the Mott insulator survives,
and each localized orbital is singly occupied by an electron, forming
a spin 1/2 magnetic moment. For stronger disorder ($W>U$) the situation
is different. Now, a fraction of electronic states are either doubly
occupied or empty, as in an Anderson insulator. The Mott gap is now
closed, although a finite fraction of the electrons still remain as
localized magnetic moments. Such a state can be described \cite{motand,vladgabisdmft2}
as an inhomogeneous mixture of a Mott and an Anderson insulator. A
very similar {}``\textbf{two-fluid model}\textquotedblright\ -
of coexisting local magnetic moments and conduction electrons - was
proposed \cite{marko,paalanen} some time ago on experimental grounds,
as a model for doped semiconductors. Some theoretical basis of such
behavior has been discussed \cite{milovanovicetal89,bhattfisher92,vladtedgabi,volfle,dk-prl93,dk-prb94,motand,vladgabisdmft2},
but the corresponding critical behavior remains a puzzle.

This physical picture of Mott is very transparent and intuitive. But
how is this strongly localized (atomic) limit approached when one
crosses the metal-insulator transition from the metallic side? To
address this question one needs a more detailed theory for the metal-insulator
transition region, which was not available when the questions posed
by Mott and Anderson were first put forward.

\subsection{Mott or Anderson... or both? }

Which of the two mechanisms dominates criticality in a given material?
This is the question often asked when interpreting experiments, but
a convincing answer is seldom given. To answer it precisely, one must
define the appropriate criteria - order parameters - characterizing
each of the two routes. The conceptually simplest theoretical framework
that introduces such order parameters is given by TMT-DMFT - which
we introduced in the preceding section, and discussed in detail in
the noninteracting limit.. As in conventional DMFT, its self-consistent
procedure formally sums-up all possible Feynman diagrams providing
local contributions to the electronic self-energy \cite{georgesrmp}.

\pagebreak
When the procedure is applied to systems with both interactions and
disorder systems, the self energy is still local, but may display
strong-site-to-site fluctuations. Its low-energy form \[
\Sigma_{i}(\omega_{n})=(1-Z_{i}^{-1})\,\omega_{n}+v_{i}-\varepsilon_{i}+\mu,\]
 defines \emph{local} Fermi liquid parameters \cite{motand,london}:
the local quasi-particle (QP) weight $Z_{i}$, and the renormalized
disorder potential $v_{i}$. This theory portrays a picture of a spatially
inhomogeneous Fermi liquid, and is able to track its evolution as
the critical point is approached.

In this language, Anderson localization, corresponding to the formation
of bound electronic states, is identified by the emergence of discrete
spectra \cite{anderson58} in the local density of states (LDOS).
As we have seen above, this corresponds \cite{motand,tmt} to the
vanishing of the \emph{typical} (geometrically averaged) LDOS $\rho_{typ}=\exp{<\ln(\rho_{i})>}$.
In contrast, Mott localization of itinerant electrons into magnetic
moments is identified by the vanishing of the local QP weights ($Z_{i}\rightarrow0$).
It is interesting and important to note that a very similar physical
picture was proposed as the key ingredient for {}``\textbf{local
quantum criticality}''\cite{sietal}, or ``\textbf{deconfined
quantum criticality}''\cite{senthiletal,senthiletal2}
at the $T=0$ magnetic ordering in certain heavy fermion systems.
A key feature in these theories is the possibility that Kondo screening
is destroyed precisely at the quantum critical point. As a result,
part of the electrons - those corresponding to tightly bound f-shells
of rare earth elements - {}``drop out\textquotedblright\ from the
Fermi surface and turn into localized magnetic moments. For this reason,
it is argued, any weak-coupling approach must fail in describing the
critical behavior. This is the mechanism several groups have attributed
to the breakdown of the Hertz-Millis theory \cite{hertz,millis} of
quantum criticality, which at present is believed to be incomplete.

Precisely the same fundamental problem clearly must be addressed for
the Mott-Anderson transition. The transmutation of a \textbf{fraction}
of electrons into local magnetic moments again can be viewed as the
suppression of Kondo screening - clearly a \textbf{non-perturbative
strong correlation effect} - that should be central to understanding
the critical behavior. To properly characterize it, one must keep
track of the evolution of the entire distribution $P(Z_{i})$ of local
quasi-particle weights - which can be directly obtained from TMT-DMFT
approach\ \cite{tmt} to the Mott-Anderson transition, which we outlined
above. The first applications of this new method to correlated systems
with disorder was carried out in recent studies by Vollhardt and collaborators
\cite{byczuk-prl05,byczuk-2009-102}, who numerically obtained the
phase diagram for the disordered Hubbard model at half-filling, and
discussed the influence of Mott-Anderson localization on magnetically
ordered phases. However, the qualitative nature of the critical behavior
in the Mott-Anderson transition in this model has not been examined
in these studies.

\subsection{\textit{\emph{Slave-boson solution}} }

In the following we use complementary semi-analytical methods supplemented
by Fermi-liquid theorems, in order to clarify the precise form of
criticality in this model \cite{aguiar-2009-102}. By making use of
scaling properties \cite{aguiar-2006-73,aguiar-2008-403} of Anderson
impurity models close to the MIT, we present a detailed analytic solution
for this problem, which emphasizes the dependence of the system properties
on its particle-hole symmetry. We consider a half-filled Hubbard model~\cite{vladgabisdmft2}
with random site energies, as given by the Hamiltonian\begin{equation}
H=-V\sum_{<ij>\sigma}c_{i\sigma}^{\dagger}c_{j\sigma}+\sum_{i\sigma}\varepsilon_{i}n_{i\sigma}+U\sum_{i}n_{i\uparrow}n_{i\downarrow}.\label{hamiltonian}\end{equation}
 Here, $c_{i\sigma}^{\dagger}$ ($c_{i\sigma}$) creates (destroys)
a conduction electron with spin $\sigma$ on site $i$, $n_{i\sigma}=c_{i\sigma}^{\dagger}c_{i\sigma}$,
$V$ is the hopping amplitude, and $U$ is the on-site repulsion.
The random on-site energies $\varepsilon_{i}$ follow a distribution
$P(\varepsilon)$, which is assumed to be uniform and have width $W$.

TMT-DMFT \cite{tmt,byczuk-prl05} maps the lattice problem onto an
ensemble of single-impurity problems, corresponding to sites with
different values of the local energy $\varepsilon_{i}$, each being
embedded in a typical effective medium which is self-consistently
calculated. In contrast to standard DMFT \cite{tanaskovicetal03},
TMT-DMFT determines this effective medium by replacing the spectrum
of the environment ({}``cavity'') for each site by its typical value,
which is determined by the process of \textit{geometric} averaging.
For a simple semi-circular model density of states, the corresponding
bath function is given by \cite{tmt,byczuk-prl05} $\Delta(\omega)=V^{2}G_{typ}(\omega)$,
with $G_{typ}(\omega)=\int_{-\infty}^{\infty}d\omega^{\prime}\rho_{typ}(\omega^{\prime})/(\omega-\omega^{\prime})$
being the Hilbert transform of the geometrically-averaged (typical)
local density of states (LDOS) $\rho_{typ}(\omega)=\exp\{\int d\varepsilon P(\varepsilon)\ln\rho(\omega,\varepsilon)\}$.
Given the bath function $\Delta(\omega)$, one first needs to solve
the local impurity models and compute the local spectra $\rho(\omega,\varepsilon)=-\pi^{-1}\operatorname{Im}G(\omega,\varepsilon)$,
and the self-consistency loop is then closed by the the geometric
averaging procedure.

To qualitatively understand the nature of the critical behavior, it
is useful to concentrate on the low-energy form for the local Green's
functions, which can be specified in terms of two Fermi liquid parameters
as \begin{equation}
G(\omega,\varepsilon_{i})=\frac{Z_{i}}{\omega-\tilde{\varepsilon}_{i}-Z_{i}\Delta(\omega)},\end{equation}
 where $Z_{i}$ is the local quasi-particle (QP) weight and $\tilde{\varepsilon}_{i}$
is the renormalized site energy \cite{tanaskovicetal03}. The parameters
$Z_{i}$ and $\tilde{\varepsilon}_{i}$ can be obtained using any
quantum impurity solver, but to gain analytical insight here we focus
on the variational calculation provided by the {}``four-boson''
technique (SB4) of Kotliar and Ruckenstein~\cite{kotliarruckenstein},
which is known to be quantitatively accurate at $T=0$. The approach
consists of determining the site-dependent parameters $e_{i}$, $d_{i}$
and $\tilde{\varepsilon}_{i}$ by the following equations \begin{equation}
-\frac{\partial Z_{i}}{\partial e_{i}}\frac{1}{\beta}\sum_{\omega_{n}}\Delta(\omega_{n})G_{i}(\omega_{n})=Z_{i}\left(\mu+\tilde{\varepsilon}_{i}-\varepsilon_{i}\right)e_{i},\end{equation}

\begin{equation}
-\frac{\partial Z_{i}}{\partial d_{i}}\frac{1}{\beta}\sum_{\omega_{n}}\Delta(\omega_{n})G_{i}(\omega_{n})=Z_{i}\left(U-\mu-\tilde{\varepsilon}_{i}+\varepsilon_{i}\right)d_{i},\end{equation}

\begin{equation}
\frac{1}{\beta}\sum_{\omega_{n}}G_{i}(\omega_{n})=\frac{1}{2}Z_{i}\left(1-e_{i}^{2}+d_{i}^{2}\right),\end{equation}
 where $Z_{i}=2(e_{i}+d_{i})^{2}[1-(e_{i}^{2}+d_{i}^{2})]/[1-(e_{i}^{2}-d_{i}^{2})^{2}]$
in terms of $e_{i}$ and $d_{i}$ and $\mu=U/2$. We should stress,
though, that most of our analytical results rely only on Fermi liquid
theorems constraining the qualitative behavior at low energy, and
thus do not suffer from possible limitations of the SB4 method.

Within this formulation, the metal is identified by nonzero QP weights
$Z_{i}$ on \textit{all} sites and, in addition, a nonzero value for
both the typical and the average {[}$\rho_{av}(\omega)=\int d\varepsilon P(\varepsilon)\rho(\omega,\varepsilon)${]}
LDOS. Mott localization (i.e. local moment formation) is signaled
by $Z_{i}\longrightarrow0$ \cite{tanaskovicetal03}, while Anderson
localization corresponds to $Z_{i}\neq0$ and $\rho_{av}\neq0$, but
$\rho_{typ}=0$ \cite{anderson58,tmt}. While Ref. \cite{byczuk-prl05}
concentrated on $\rho_{typ}$ and $\rho_{av}$, we find it useful
to simultaneously examine the QP weights $Z_{i}$, in order to provide
a complete and precise description of the critical behavior.

\subsection{\textit{\emph{Phase diagram}} }

\begin{figure}[h]
 \vspace{24pt}
\begin{centering}

\includegraphics[width=10cm]{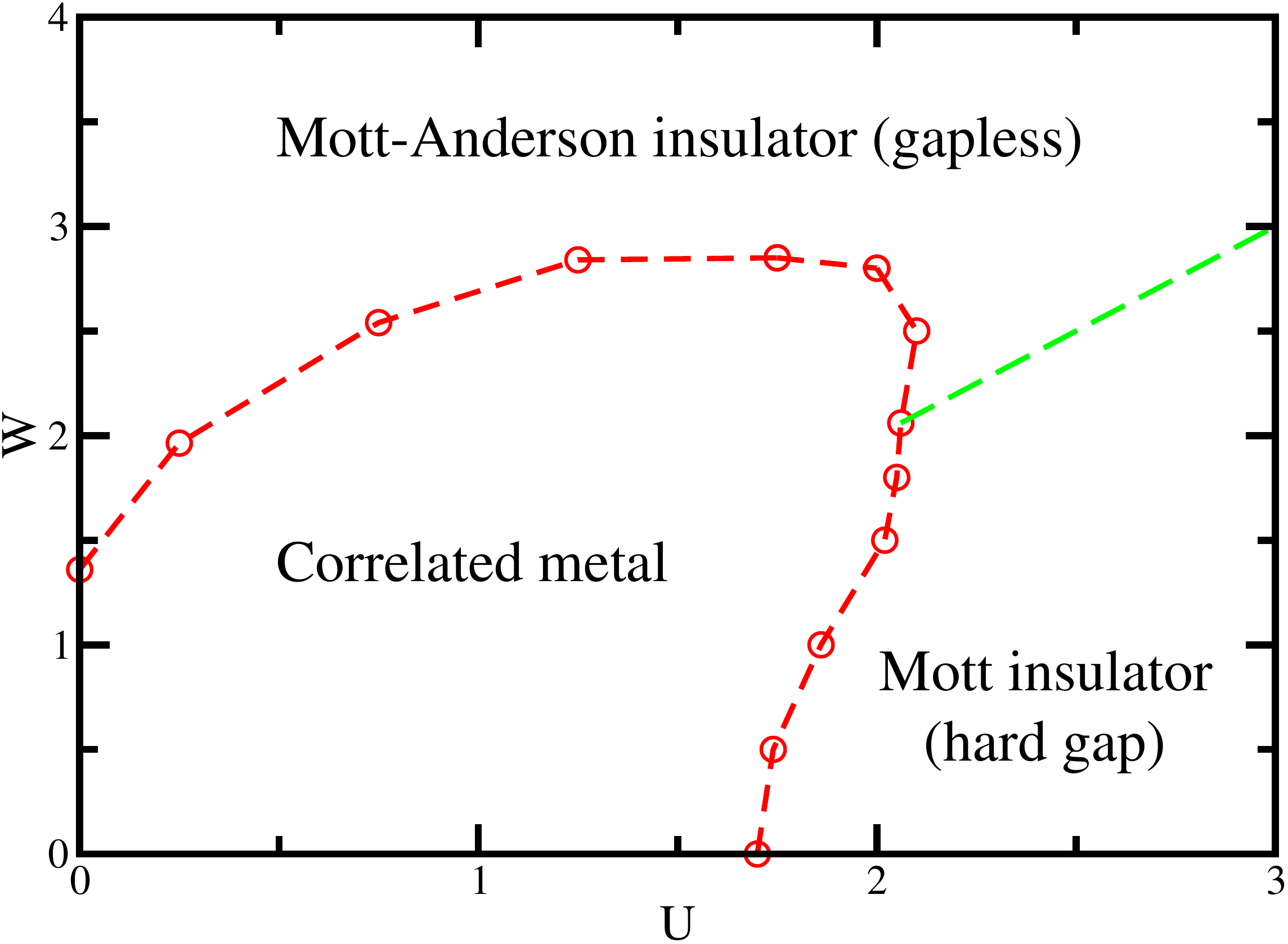} 
\par\end{centering}

\caption{$T=0$ phase diagram for the disordered half filled
Hubbard model, obtained from the numerical SB4 solution of TMT-DMFT. }

\end{figure}

Using our SB4 method, the TMT-DMFT equations can be numerically solved
to very high accuracy, allowing very precise characterization of the
critical behavior. In presenting all numerical results we use units
such that the bandwidth $B=4V=1$. Fig. 16 shows the resulting $T=0$
phase diagram at half filling, which generally~agrees with that of
Ref.~\cite{byczuk-prl05}. By concentrating first on the critical
behavior of the QP weights $Z_{i}$, we are able to clearly and precisely
distinguish the metal from the insulator. We find that at least some
of the $Z_{i}$ vanish all along the phase boundary. By taking a closer
look, however, we can distinguish two types of critical behavior,
as follows.

\subsubsection{\emph{Mott-Anderson vs. Mott-like transition} }

For sufficiently strong disorder ($W>U$), the Mott-Anderson transition
proves qualitatively different than the clean Mott transition, as
seen by examining the critical behavior of the QP weights $Z_{i}=Z(\varepsilon_{i})$.
Here $Z_{i}\rightarrow0$ only for $0<|\varepsilon_{i}|<U/2$ , indicating
that only \textit{a fraction} of the electrons turn into localized
magnetic moments. The rest show $Z_{i}\rightarrow1$ and undergo Anderson
localization (see below). Physically, this regime corresponds to a
spatially inhomogeneous system, with Mott fluid droplets interlaced
with regions containing Anderson-localized quasiparticles. In contrast,
for weaker disorder ($W<U$) the transition retains the conventional
Mott character. In this regime $Z_{i}\rightarrow0$ on all sites,
corresponding to Mott localization of all electrons. We do not discuss
the coexistence region found in Ref.~\cite{byczuk-prl05}, because
we focus on criticality within the metallic phase. We do not find
any {}``crossover'' regime such as reported in Ref.~\cite{byczuk-prl05}
the existence of which we believe is inconsistent with the generally
sharp distinction between a metal and an insulator at $T=0$.

\subsubsection{\emph{Two-fluid behavior at the Mott-Anderson transition}}

To get a closer look at the critical behavior of the QP weights $Z_{i}=Z(\varepsilon_{i})$,
we monitor their behavior near the transition. The behavior of these
QP weights is essentially controlled by the spectral weight of our
self-consistently-determined TMT bath, which we find to vanish at
the transition. An appropriate parameter to measure the distance to
the transition is the bandwidth $t$ of the bath spectral function,
which is shown in Fig. 17.

\begin{figure}[h]
 
\begin{centering}
\includegraphics[width=12cm]{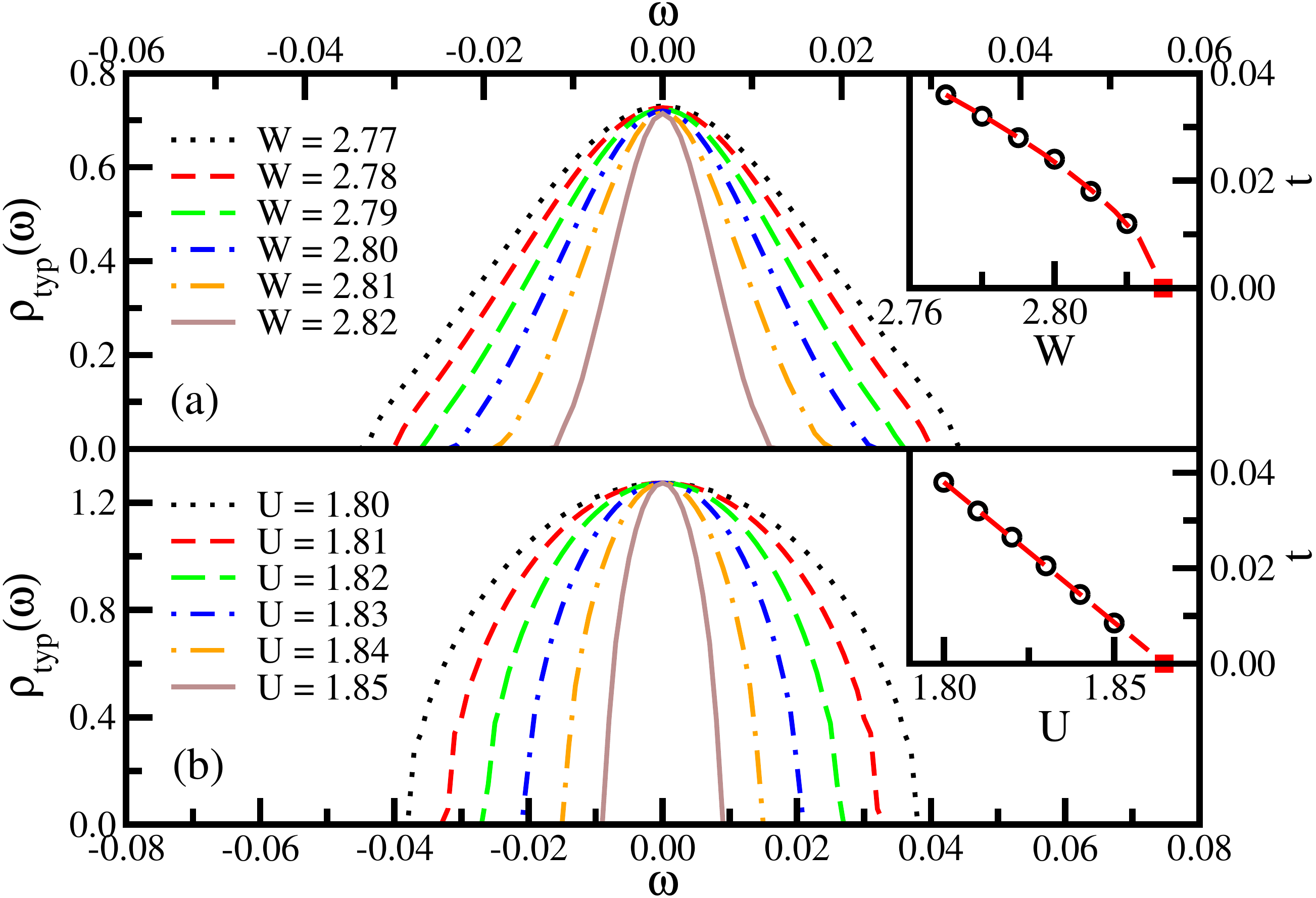} 
\par\end{centering}

\caption{Frequency dependence of the typical DOS very close
to the metal-insulator transition for (a) the Mott-Anderson transition
($W>U$) at $U=1.25$ and (b) the Mott-like transition ($W<U$) at
$W=1.0$. The insets show how, in both cases, the $\rho_{typ}(\omega)$
bandwidth $t\rightarrow0$ at the transitions.}

\end{figure}

Considering many single-impurity problems, we observe a two-fluid
picture, just as in the limit earlier analyzed by Mott.~\cite{mott-book90}
Indeed, these results correspond to the same atomic limit discussed
by Mott, since, although the hopping itself is still finite, the cavity
field {}``seen'' by the impurities goes to zero in the current case.

As in the atomic limit, the sites with $|\varepsilon_{i}|<U/2$ turn
into local moments and have vanishing quasiparticle weight $Z_{i}\rightarrow0$.
The remaining sites show $Z_{i}\rightarrow1$, as they are either
doubly occupied, which corresponds to those with $\varepsilon_{i}<-U/2$,
or empty, which is the case for those sites with $\varepsilon_{i}>U/2$.
Consequently, as the transition is approached, the curves $Z(\varepsilon_{i},t)$
{}``diverge'' and approach either $Z=0$ or $Z=1$. These values
of $Z$ can thus be identified as two stable fixed points for the
problem in question, as we discuss below.
\pagebreak

\begin{figure}[h]

\begin{centering}
\includegraphics[width=11cm]{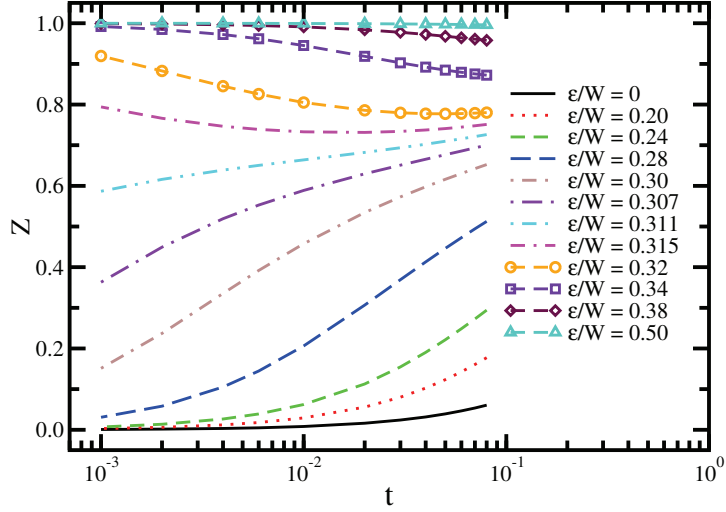} 
\par\end{centering}

\caption{Quasiparticle weight $Z$ plotted as a function of
the distance to the Mott-Anderson transition $t$, for different values
of the local site energy $\epsilon/W$. We present the results only
for positive site energies, as a similar behavior holds for negative
ones. }

\label{fig8} 
\end{figure}

Note that in Fig.~\ref{fig8} we restrict the results to positive
energy values, as a similar behavior is observed for negative $\varepsilon_{i}$.
In this case, there is precisely one value of the site energy $\varepsilon_{i}=\varepsilon^{\ast}$,
for which $Z(\varepsilon^{\ast},t)\rightarrow Z^{\ast}$. This corresponds
to the value of $\varepsilon_{i}$ below which $Z$ {}``flows''
to $0$ and above which $Z$ {}``flows'' to $1$. In other words,
it corresponds to an unstable fixed point. Just as in the atomic limit,
$\varepsilon^{\ast}$ is equal to $U/2$ ($\varepsilon^{\ast}/W=0.3125$
in Fig.~\ref{fig8}, where $U=1.75$ and $W=2.8$).

\begin{figure}[h]

\begin{centering}
\includegraphics[width=11cm]{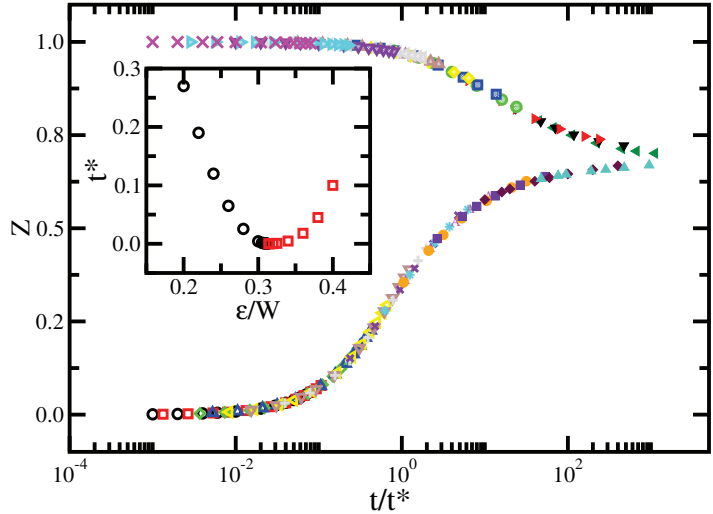} 
\par\end{centering}
\caption{Quasiparticle weight $Z$ as a function of $t/t^{\ast}(\delta\varepsilon)$
showing that the results for different $\varepsilon$ can be collapsed
onto a single scaling function with two branches. The results for
different $\varepsilon$ correspond to different symbols. The inset
shows the scaling parameter $t^{\ast}$ as a function of $\varepsilon/W$
for the upper (squares) and bottom (circles) branches. }
\label{fig2} 
\end{figure}

\subsubsection{$\beta$\emph{-function formulation of scaling}}

Our numerical solutions provide evidence that as a function of $t$
the {}``charge'' $Z(t)$ {}``flows'' away from the unstable {}``fixed
point'' $Z^{\ast}$, and towards either stable {}``fixed points''
$Z=0$ or $Z=1$. The structure of these flows show power-law scaling
as the scale $t\rightarrow0$; this suggests that it should be possible
to collapse the entire family of curves $Z(t,\delta\varepsilon)$
onto a single universal scaling function \begin{equation}
Z(t,\delta\varepsilon)=f[t/t^{\ast}(\delta\varepsilon)],\label{scaling}\end{equation}
 where the crossover scale $t^{\ast}(\delta\varepsilon)=C^{\pm}|\delta\varepsilon|^{\phi}$
around the unstable fixed point. Remarkably, we have been able to
scale the numerical data precisely in this fashion, see Fig.~\ref{fig2},
and extract the form of $t^{\ast}(\delta\varepsilon)$. We find that
$t^{\ast}(\delta\varepsilon)$ vanishes in a power law fashion at
$\delta\varepsilon=0$, with exponent $\phi=2$ and the amplitudes
$C^{\pm}$ differ by a factor close to two for $Z\gtrless Z^{*}$.

As shown in Fig.~\ref{fig2}, the scaling function $f(x)$ where
$x=t/t^{\ast}(\delta\varepsilon)$ presents two branches: one for
$\varepsilon_{i}<\varepsilon^{\ast}$ and other for $\varepsilon_{i}>\varepsilon^{\ast}$.
We found that for $x\rightarrow0$ both branches of $f(x)$ are linear
in $x$, while for $x\gg1$ they merge, i.e. $f(x)\rightarrow Z^{\ast}\pm A^{\pm}x^{-1/2}$.
As can be seen in the first two panels, in the limit $t\rightarrow0$,
the curve corresponding to $\varepsilon_{i}<U/2$ has $Z(t)=B^{-}t$,
while that for $\varepsilon_{i}>U/2$ follows $1-Z(t)=B^{+}t$. These
results are for a flat cavity field but, as mentioned earlier, we
checked that the same exponents are found also for other bath functions,
meaning that they are independent of the exact form of the cavity
field. The power-law behavior and the respective exponents observed
numerically in the three limits above have also been confirmed by
solving the SB equations analytically \cite{aguiar-2008-403}close
to the transition ($t\rightarrow0$).

In the following, we rationalize these findings by defining an appropriate
$\beta$-function which describes all the fixed points and the corresponding
crossover behavior. Let us assume that \begin{equation}
\frac{dZ(t,\delta\varepsilon)}{d\ln t}=-\beta(Z)\label{defbeta}\end{equation}
 is an explicit function of $Z$ only, but not of the parameters $t$
or $\delta\varepsilon$. The desired structure of the flows would
be obtained if the $\beta$-function had three zeros: at $Z=0$ and
$Z=1$ with negative slope (stable fixed points) and one at $Z=Z^{\ast}$
with positive slope (unstable fixed point). The general structure
of these flows can thus be described in a $\beta$-function language
similar to that used in the context of a renormalization group approach;
we outline the procedure to obtain $\beta(Z)$ from the numerical
data.

\begin{figure}[h]
 
\begin{centering}
\includegraphics[width=11cm]{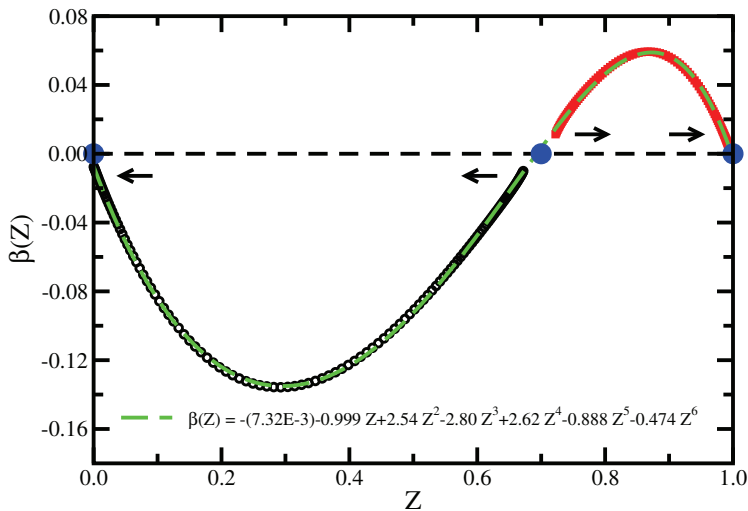} 
\par\end{centering}

\caption{$\beta$-function obtained as described in the text
for the Anderson impurity models close to the Mott-Anderson transition.
The filled circles indicate the three fixed points found for this
problem. The arrows indicate how $Z$ flows \textit{to} the stable
points ($Z=0$ and $Z=1$) and \textit{from} the unstable one ($Z\approx0.7$).}

\label{fig6} 
\end{figure}

The integration of Eq.~(\ref{defbeta}) can be written in the form
of Eq.~(\ref{scaling}) as \begin{equation}
Z=f[t/t^{\ast}(Z_{o})],\end{equation}
 where $Z_{o}$ is the initial condition (a function of $\delta\varepsilon$).
With $x=t/t^{*}$ as before, Eq.~(\ref{scaling}) can be rewritten
as 
\begin{equation}
\beta(Z)=-xf^{\prime}(x).\label{beta}
\end{equation}

 The numerical data for $Z=f(x)$ as a function of $x$ is presented
in Fig.~\ref{fig2}. Thus, using Eq.~(\ref{beta}), the $\beta$-function
in terms of $x(Z)$ is determined, which can finally be rewritten
in terms of $Z$. Carrying out this procedure, we obtain $\beta(Z)$
as shown in Fig.~\ref{fig6}. In accordance with what was discussed
above, we see that $\beta(Z)$ has three fixed points, as indicated
in the figure by filled circles. $Z=0$ and $Z=1$ are stable, while
$Z\approx0.7$ is the unstable fixed point. The scaling behavior and the associated $\beta$-function observed here reflect the fact these impurity models have two phases (singlet and doublet) when entering the insulator. The two stable fixed points
describe these two phases, while the unstable fixed point $Z^{\ast}$
describes the phase transition, which is reached by tuning the site
energy. 

Interestingly, the family of curves in Fig.~\ref{fig8} looks similar
to those seen in some other examples of quantum critical phenomena.
In fact, one can say that the crossover scale $t$ plays the role
of the reduced temperature, and the reduced site energy $\delta\varepsilon=\left(\varepsilon_{i}-\varepsilon^{\ast}\right)/\varepsilon^{\ast}$
that of the control parameter of the quantum critical point. As the
site energy is tuned at $t=0$, the impurity model undergoes a phase
transition from a singlet to a doublet ground state. Quantum fluctuations
associated with the metallic host introduce a cutoff and round this
phase transition, which becomes sharp only in the $t\rightarrow0$
limit.

\subsection{\textit{\emph{Wavefunction localization}} }

To more precisely characterize the critical behavior we now turn our
attention to the spatial fluctuations of the quasiparticle wavefunctions,
we compare the behavior of the typical ($\rho_{typ}$) and the average
($\rho_{av}$) LDOS. The approach to the Mott-Anderson transition
($W>U$) is illustrated by increasing disorder $W$ for fixed $U=1.25$
(Fig. 21 - top panels). Only those states within a narrow energy range
($\omega<t$, see also Fig. 17) around the band center (the Fermi
energy) remain spatially delocalized ($\rho_{typ}\sim\rho_{av}$),
due to strong disorder screening \cite{tanaskovicetal03,aguiar-2008-403}
within the Mott fluid (sites showing $Z_{i}\rightarrow0$ at the transition).
The electronic states away from the band center (i.e. in the band
tails) quickly get Anderson-localized, displaying large spatial fluctuations
of the wavefunction amplitudes \cite{tmt} and having $\rho_{typ}\ll\rho_{av}$.

\begin{figure}[h]

\begin{centering}
\includegraphics[width=11cm]{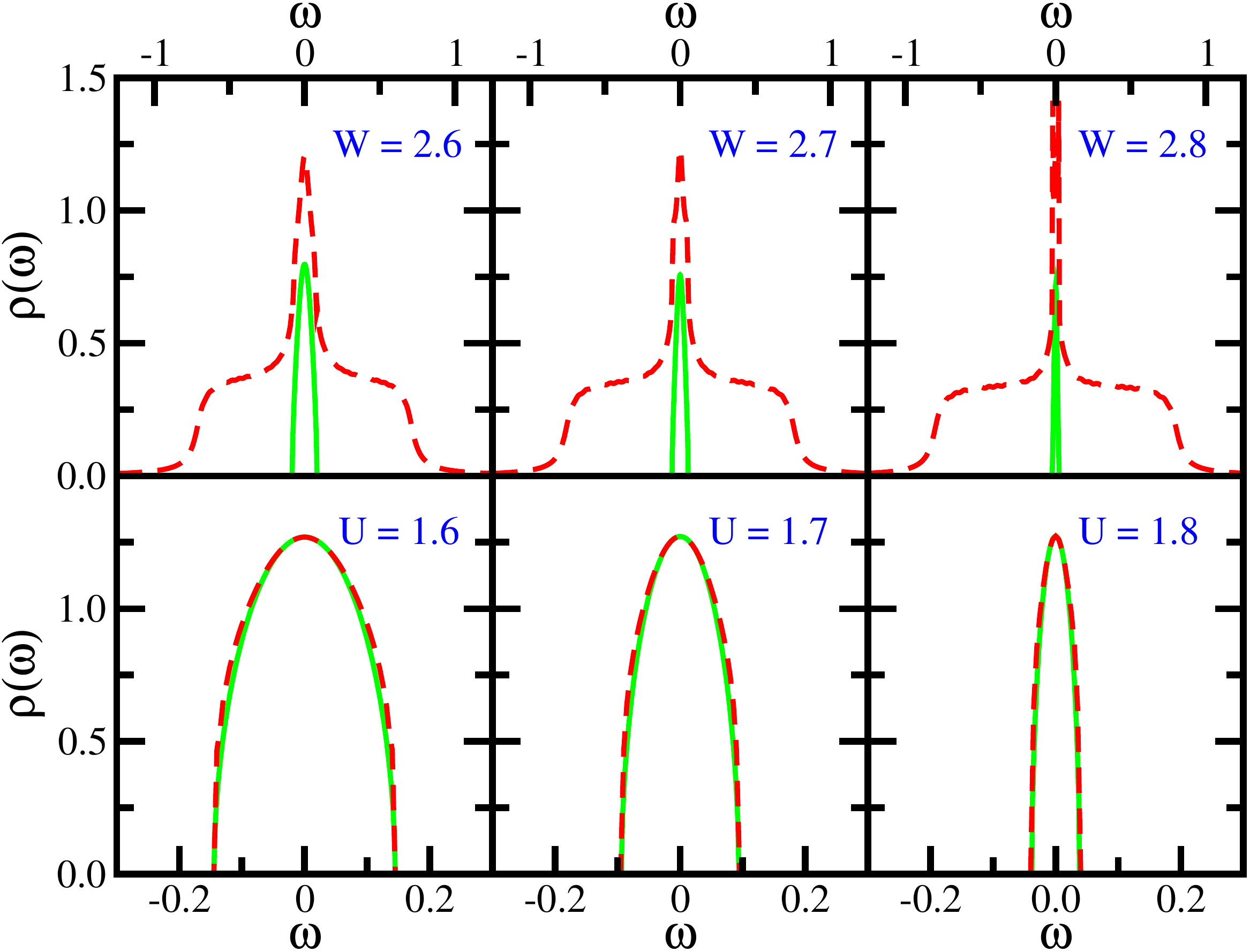} 
\par\end{centering}

\caption{Frequency dependence of $\rho_{typ}$ (full line) and
$\rho_{av}$ (dashed line) in the critical region. Results in top
panels illustrate the approach to the Mott-Anderson transition ($W>U$)
at $U=1.25$; the bottom panels correspond to the Mott-like transition
($W<U$) at $W=1.0$.  For the Mott-Anderson transition, only a narrow band of delocalized states remain near the Fermi energy, corresponding to  $\rho_{typ}\neq 0$. In contrast, most electronic states remain delocalized  $\rho_{typ}\approx \rho_{av}$ near the Mott-like transition.}

\end{figure}

The spectral weight of the delocalized states (states in the range
$\omega<t$) decreases with disorder and vanishes at the transition,
indicating the Mott localization of this fraction of electrons. At
this critical point, the crossover scale $t$ also vanishes. In contrast,
the \textit{height} $\rho_{typ}(0)$ remains finite at the transition,
albeit at a reduced $W$-dependent value, as compared to the clean
limit. More precise evolution of $\rho_{typ}(0)$ is shown in Fig.~22a,
demonstrating its critical jump.

Behavior at the Mott-like transition ($W<U$) is dramatically different
(Fig.~21 - bottom panel). Here $\rho_{typ}\approx\rho_{av}$ over
the entire QP band, indicating the absence of Anderson localization.
It proves essentially identical as that established for the disordered
Hubbard model within standard DMFT~\cite{tanaskovicetal03}, reflecting
strong correlation-enhanced screening of disorder \cite{tanaskovicetal03,aguiar-2008-403},
where both $\rho_{av}(\omega=0)$ and $\rho_{typ}(\omega=0)$ approach
the bare ($W=0$) value (see also Fig.~22b). Similar results were
found in Ref. \cite{byczuk-prl05}, but an explanation was not provided.

The corresponding pinning \cite{tanaskovicetal03,aguiar-2008-403}
for $\rho(\omega=0,\varepsilon)$ is shown in the insets of Fig.~3,
both for the Mott-Anderson and the Mott-like transition. In the Mott-Anderson
case, this mechanism applies only within the Mott fluid ($|\varepsilon|<U/2$),
while within the Anderson fluid ($|\varepsilon|>U/2$) it assumes
smaller values, explaining the reduction of $\rho_{typ}(0)$ in this
case.

\subsection{\textit{\emph{Analytical solution}} }

Within our SB4 approach, the TMT-DMFT order-parameter function $\rho_{typ}(\omega)$
satisfies the following self-consistency condition\begin{align}
\rho_{typ}(\omega) & =\exp\int d\varepsilon P\left(\varepsilon\right)\left\{ \ln[V^{2}Z^{2}(\varepsilon)\rho_{typ}(\omega)]\right.\nonumber \\
 & -\ln[(\omega-\widetilde{\varepsilon}(\varepsilon)-V^{2}Z(\varepsilon)\operatorname{Re}G_{typ}(\omega))^{2}\nonumber \\
 & \left.+(\pi V^{2}Z(\varepsilon)\rho_{typ}(\omega))^{2}]\right\} .\end{align}
 While the solution of this equation is in general difficult, it simplifies
in the critical region, where the QP parameter functions $Z(\varepsilon)$
and $\widetilde{\varepsilon}(\varepsilon)$ assume scaling forms which
we carefully studied in previous work \cite{aguiar-2008-403}. This
simplification allows, in principle, to obtain a closed solution for
all quantities. In particular, the crossover scale $t$, which defines
the $\rho_{typ}(\omega)$ mobility edge (see Fig.~17 and Ref.~\cite{aguiar-2008-403}),
is determined by setting $\rho_{typ}(\omega=t)=0$.

Using this approach we obtain that, in the case of Mott-like transition
($W<U$), the critical behavior of all quantities reduces to that
found in standard DMFT~\cite{tanaskovicetal03}, including $t\sim U_{c}(W)-U$
(in agreement with the numerical results of Fig.~17b), perfect screening
of site randomness~\cite{tanaskovicetal03,aguiar-2008-403}, and
the approach of $\rho_{av}(\omega=0)$ and $\rho_{typ}(\omega=0)$
to the clean value. The precise form of the critical behavior for
the crossover scale $t$ is more complicated for the Mott-Anderson
transition ($W>U$) (as confirmed by our numerical results in Fig.~17a),
and this will not be discussed here.%
\begin{figure}[h]

\begin{centering}
\includegraphics[width=12cm]{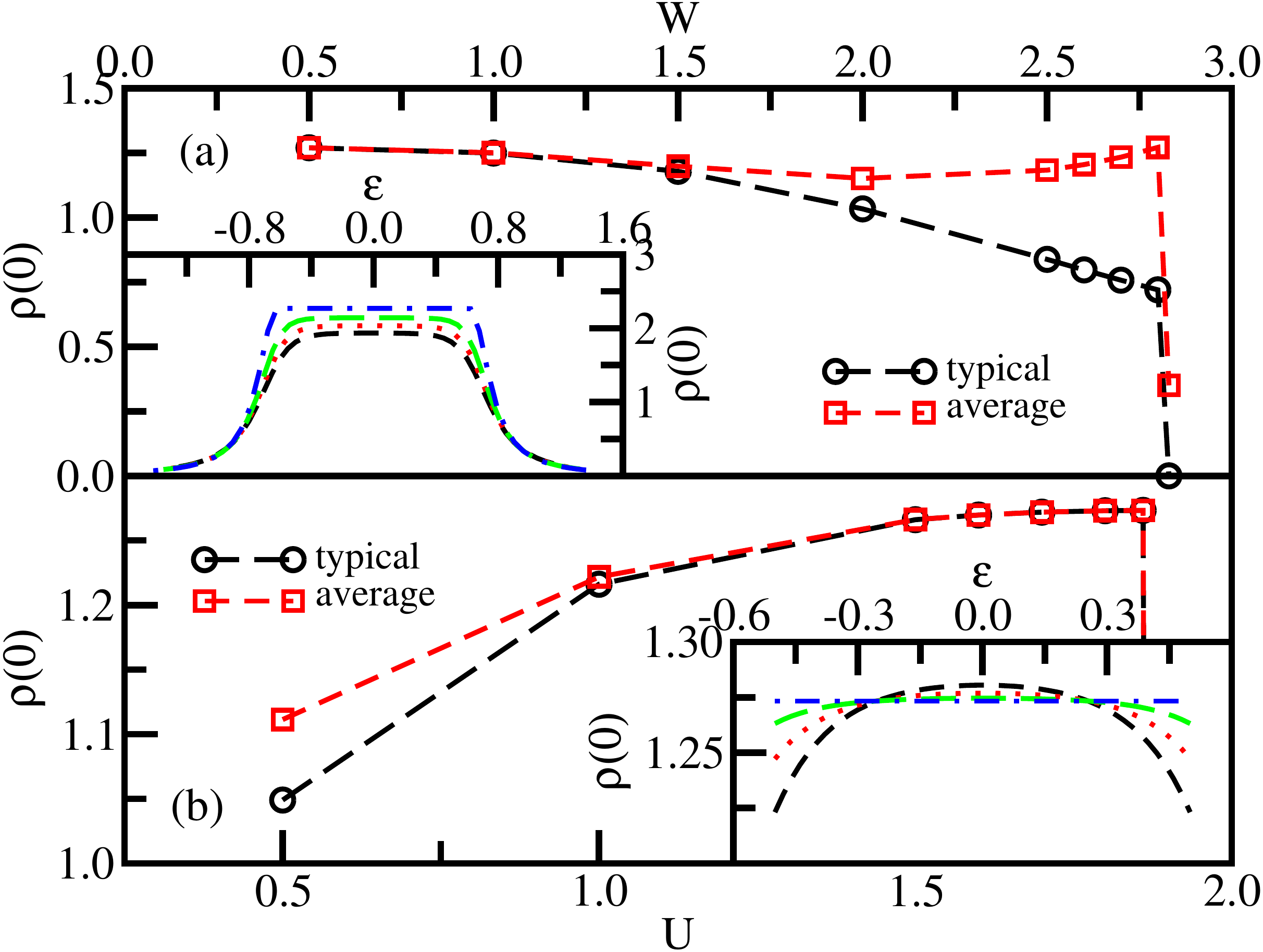} 
\par\end{centering}

\caption{Typical and average values of $\rho(0)$ as the metal-insulator
transition is approached for (a)~$U=1.25$ and (b) $W=1.0$. The
insets show $\rho(0)$ as a function of $\varepsilon$ for (a) $W=2.5$,
$2.6$, $2.7$ and $2.83$ (from the black curve to the blue one)
and (b) $U=1.5$, $1.6$, $1.7$ and $1.86$.}

\end{figure}

Instead, we focus on elucidating the origin of the puzzling behavior
of $\rho_{c}=\rho_{typ}(\omega=0)$, which is known \cite{tmt} to
vanish linearly $\rho_{c}\sim(W_{c}-W)$ for $U=0$, but which we
numerically find to display a jump (i.e. a finite value) at criticality,
as soon as interactions are turned on. For $\omega=0$ our self-consistency
condition reduces (for our model $\mbox{Re}G_{typ}(0)=0$ by particle-hole
symmetry) to \begin{equation}
\int d\varepsilon P\left(\varepsilon\right)\ln\frac{V^{2}Z^{2}(\varepsilon)}{\widetilde{\varepsilon}(\varepsilon)^{2}+\pi^{2}V^{4}Z^{2}(\varepsilon)\rho_{c}^{2}}=0,\label{omega0}\end{equation}
 which further simplifies as we approach the critical point. Here,
the QP parameters $Z(\varepsilon)\longrightarrow0$ and $\widetilde{\varepsilon}(\varepsilon)\sim$
$Z^{2}(\varepsilon)\ll$ $Z(\varepsilon)$ for the Mott fluid ($|\varepsilon|<U/2$),
while $Z(\varepsilon)\longrightarrow1$ and $|\widetilde{\varepsilon}(\varepsilon)|\longrightarrow$
$|\varepsilon-U/2|$ for the Anderson fluid ($|\varepsilon|>U/2$),
and we can write\begin{align}
0 & =\int_{0}^{U/2}\hspace{-12pt}d\varepsilon P\left(\varepsilon\right)\ln\frac{1}{\left(\pi V\rho_{c}\right)^{2}}\nonumber \\
 & -\int_{0}^{(W-U)/2}\hspace{-32pt}d\varepsilon P\left(\varepsilon\right)\ln[\left(\varepsilon/V\right)^{2}+\left(\pi V\rho_{c}\right)^{2}].\label{rhoc}\end{align}

This expression becomes even simpler in the $U<<W$ limit, giving
\begin{equation}
\frac{U}{W}\ln\frac{1}{\pi V\rho_{c}}+a-bV\rho_{c}+O[\rho_{c}^{2}]=0,\label{smallU}\end{equation}
 where $a(W,U)=(1-U/W)\{1-\ln[(W-U)/2V]\}$ and $b=\frac{2\pi^{2}V}{W}$.
This result reproduces the known result \cite{tmt} $\rho_{c}\sim(W_{c}-W)$
at $U=0$, but dramatically different behavior is found as soon as
$U>0$. Here, a \textit{non-analytic} (singular) contribution emerges
from the Mott fluid ($|\varepsilon|<U/2$), which assures that $\rho_{c}$
must remain finite at the critical point, consistent with our numerical
results (see Fig.~22). Note that the second term in Eq.~(\ref{rhoc}),
coming from the Anderson fluid ($|\varepsilon|>U/2$), vanishes in
the case of a Mott-like transition ($U>W$), and our result reproduces
the standard condition $\pi\rho_{c}V=1$~\cite{tanaskovicetal03},
which corresponds to the clean limit.

A further glimpse on how the condition $\pi\rho_{c}V=1$ is gradually
violated as we cross on the Mott-Anderson side is provided by solving
Eq.~(\ref{rhoc}) for $U\precsim W$ limit, giving \begin{equation}
\rho_{c}\approx\frac{1}{\pi V}\left[1-\frac{1}{24}\left(\frac{W}{V}\right)^{2}\left(1-\frac{U}{W}\right)^{3}\right],\end{equation}
 again consistent with our numerical solution.

But what is the physical origin of the jump in $\rho_{c}$? To see
it, note that the singular form of the first term in Eq.~(\ref{rhoc})
comes from the Kondo pinning\cite{tanaskovicetal03} $\widetilde{\varepsilon}(\varepsilon)\sim$
$Z^{2}(\varepsilon)\ll$ $Z(\varepsilon)$ within the Mott fluid.
This behavior reflects the particle-hole symmetry of our (geometrically
averaged) $\rho_{typ}(\omega=0)$ bath function, which neglects site-to-site
cavity fluctuations present, for example, in more accurate statDMFT
theories \cite{motand,london,mirandavlad1,andrade09prl,andrade09physicsB,miranda-dobrosavljevic-rpp05}.
Indeed, in absence of particle-hole symmetry, one expects \cite{tanaskovicetal03}
$\widetilde{\varepsilon}(\varepsilon)\sim$ $Z(\varepsilon)$, and
the resulting $\varepsilon$-dependence should cut-off the log singularity
responsible for the jump in $\rho_{c}$. This observation provides
a direct path to further refine the TMT-DMFT approach, reconciling
the present results with previous statDMFT findings \cite{motand,london,mirandavlad1,andrade09prl,andrade09physicsB,miranda-dobrosavljevic-rpp05}.
As a next step, one should apply the TMT ideas to appropriately chosen
effective models~\cite{tanaskovicetal04}, in order to eliminate
those features reflecting the unrealistic particle-hole symmetry built
in the current theory. We emphasize that the two-fluid picture is
a consequence of only a fraction of the sites showing $Z\rightarrow0$
and is not dependent on either particle-hole symmetry or the consequent
jump in the DOS. This interesting research direction is just one of 
many possible future applications of our TMT-DMFT formalism.

\pagebreak
\section{\textit{\emph{Conclusions and outlook}} }

This article described the conceptually simplest theoretical approach
which is able to capture the interplay of strong correlation effects
- the Mott physics - and the disorder effects associated with Anderson
localization. It demonstrated that one can identify the signatures
of both of these basic mechanisms for localization by introducing
appropriate \emph{local order parameters,} which are then self-consistently
calculated within the proposed \textbf{Typical-Medium Theory}. We
showed that key insight can be obtained by focusing on the evolution
of the local quasiparticle weights $Z_{i}$ as a \textit{second order
parameter} describing tendency to Mott localization, in addition to
the Anderson-like TMT order parameter $\rho_{typ}$. Our main finding
is that, for sufficiently strong disorder, the physical mechanism
behind the Mott-Anderson transition is the formation of two fluids,
a behavior that is surprisingly reminiscent of the phenomenology proposed
for doped semiconductors \cite{paalanenetal88}. Here, only a fraction
of the electrons (sites) undergo Mott localization; the rest can be
described as Anderson-localized quasiparticles. Physically, it describes
spatially inhomogeneous situations, where the Fermi liquid quasiparticles
are destroyed only in certain regions - the Mott droplets - but remain
coherent elsewhere. Thus, in our picture the Mott-Anderson transition
can be seen as reminiscent of the {}``orbitally selective'' Mott
localization~\cite{deleo-2008-101,pepin-2007-98}. To be more precise,
here we have a {}``site selective'' Mott transition, since it emerges
in a spatially resolved fashion. Understanding the details of such
{}``site selective'' Mott transitions should be viewed as an indispensable
first step in solving the long-standing problem of metal-insulator
transitions in disordered correlated systems.

\section*{Acknowledgements}
The author thanks Elihu Abrahams, Carol Aguiar, Eric Andrade, Gabi
Kotliar, Eduardo Miranda, Andrei Pastor, and Darko Tanaskovi\'{c}
for many years of exciting and fruitful collaboration. This work was
supported by the NSF grant DMR-0542026.

\bibliographystyle{ws-rv-van}

\end{document}